\documentclass[aps,pra,letterpaper,twocolumn,superscriptaddress,groupedaddress]{revtex4}

\usepackage{amssymb}
\usepackage{color,graphicx}
\usepackage{amsmath}
\usepackage{amsbsy}
\usepackage{amsthm}
\usepackage{bbm}
\usepackage{bm}
\usepackage{epsfig}
\usepackage{lscape}
\usepackage{float}
\usepackage{subfigure}
\usepackage{multirow}

\def\be{\begin{equation}}
\def\ee{\end{equation}}

\def\CNOT{\mathop{\tt CNOT}\nolimits}

\begin{document}
\title{Bacon-Shor code with continuous measurement of non-commuting operators}

\author{Juan Atalaya}
\thanks{These authors contributed equally to this work}
\affiliation{Department of Electrical and Computer Engineering, University of California, Riverside, California 92521, USA}
\author{Mohammad Bahrami}
\thanks{These authors contributed equally to this work}
\affiliation{Department of Electrical and Computer Engineering, University of California, Riverside, California 92521, USA}
\author{Leonid P. Pryadko}
\affiliation{Department of Physics \& Astronomy, University of
  California, Riverside, California 92521, USA}
\author{Alexander N. Korotkov}
\affiliation{Department of Electrical and Computer Engineering, University of California, Riverside, California 92521, USA}

\date{\today}

\begin{abstract}
We analyze the operation of a four-qubit Bacon-Shor code with simultaneous continuous measurement of non-commuting gauge operators. The error syndrome in this case is monitored via time-averaged cross-correlators of  the output signals.
We find the logical error rate for several models of decoherence, and also find the termination  rate for this quantum error detecting  code. The code operation is comparable to that based on projective measurements when the collapse timescale due to continuous measurements is an order of magnitude less than the time period between the projective measurements. An advantage of the continuous-measurement implementation is the absence of time-dependence in the code operation, with passive continuous monitoring of the error syndrome.
\end{abstract}

\maketitle


\section{Introduction}

Quantum error correction (QEC) is a necessary procedure in a practical quantum computer operation \cite{Shor-1995,Steane-1996,Gottesman-1996,N-C-book}. Besides standard stabilizer codes \cite{Gottesman-1996}, much attention has been recently given to surface codes \cite{Bravyi-1998,Raussendorf-2007,Fowler-2012} because of their relatively high fault-tolerant threshold without the need of concatenation, and also because the measured operators involve only four neighboring qubits. Significant attention has also been paid recently to Bacon-Shor codes \cite{Poulin-2005,Bacon-2006,Terhal-2015}, where measured operators involve only two qubits, which simplifies implementation. There has been a significant experimental progress toward practical QEC  \cite{Cory-1998,Chiaverini-2004,Schindler-2011,Nigg-2014,Waldherr-2014,Cramer-2016}, including experiments with superconducting qubits \cite{Reed-2012,Zhong-2014,Kelly-2015,Corcoles-2015,Riste-2015,Ofek-2016}.

While most of the QEC codes are based on repetitive projective measurement of multi-qubit operators, continuous QEC has also been analyzed theoretically \cite{Ahn-2002, Ahn-2003, Ahn-2004, Sarovar-2004,vanHandel-2005,Sarovar-2005,Oreshkov-2007,Chase-2008,Hsu-2016}. The general idea in most of these proposals is to monitor multi-qubit operators continuously and apply a continuously changing feedback Hamiltonian to the qubits. It is expected that such continuous error correction can outperform traditional QEC; however, there are significant challenges, including computationally expensive tracking of the state and the fact that the feedback Hamiltonian necessarily contains fluctuations caused by the output noise of continuous detectors.
Therefore, while continuous quantum feedback \cite{Wiseman-1993,Ruskov-2002} is already available for superconducting qubits \cite{Vijay-2012,deLange-2014}, it is still unclear in which manner it can be useful for practical QEC.

A natural way of employing continuous measurement in stabilizer codes is using it only for continuous monitoring of the error syndrome, while error correction is still applied in a traditional discrete way after the syndrome indicates that a certain error has occurred (the actual error correction can be postponed until the end of the procedure, after tracked accumulation of several errors \cite{Knill-2005,vanHandel-2005,Terhal-2015}; performance comparison with the continuous feedback is still an important subject).  This manner of operation can also be  applied to surface codes \cite{Fowler-2012} in principle, since all measured operators commute with each other. However, it is not immediately clear if continuous measurement can or cannot be used in the Bacon-Shor codes, which necessarily need measurement of non-commuting two-qubit operators. This is the question, which we analyze in this paper for the simplest four-qubit Bacon-Shor code.

Simultaneous measurement of non-commuting observables has been discussed long ago \cite{Arthurs-1965,She-1966,Helstrom-74, Yuen-91,Schroeck-1982, Busch-1985, Stenholm-1992, Ozawa-2004, Jordan-2005, Wei-2008}; however, a theory for the qubit evolution due to  continuous  non-commuting measurements has been developed relatively recently \cite{Ruskov-2010}, and the first such experiment with a superconducting qubit has been realized only in the past year \cite{Hacohen-Courgy-2016}. Note that in this experiment the physical qubit was under constant Rabi rotation, so that simultaneous measurement of non-commuting observables was realized for an effective qubit in the rotating frame. There are no experiments yet on simultaneous continuous measurement of non-commuting two-qubit operators; however, qubit entanglement due to continuous measurement of two-qubit operators \cite{Ruskov-2003} has already been well demonstrated with superconducting qubits in various setups \cite{Riste-2013,Roch-2014}. In this paper we assume simultaneous continuous measurement of non-commuting two-qubit operators without discussing possible experimental ways of realizing such a measurement (which may rely on the rotating frame as in Ref.\ \cite{Hacohen-Courgy-2016}).

The main question of this paper is whether and how continuous measurement can be used in the operation of the Bacon-Shor code, which by construction relies on non-commuting two-qubit operators. We will consider the simplest Bacon-Shor code, which contains four qubits and needs measurement of four (gauge) operators: $X_1 X_2$, $X_3X_4$, $Z_1Z_3$, and $Z_2Z_4$ (out of six pairs of these operators, four are non-commuting). The standard operation cycle of this code consists of two steps: simultaneous projective measurement of commuting operator pairs $X_1 X_2$ and $X_3X_4$, and then the second pair: $Z_1Z_3$ and $Z_2Z_4$. In contrast, in our case all four operators are measured at the same time continuously. The error syndrome in this case is monitored using time-averaged cross-correlators of the noisy output signals, so that an error is indicated by crossing a certain threshold. It is interesting that the evolution analysis is similar to the analysis of continuous non-commuting measurement of a single qubit \cite{Ruskov-2010,Hacohen-Courgy-2016,Atalaya-2016}.
Our main result is that the operation of the four-qubit Bacon-Shor code with continuous measurement is indeed possible and similar to the standard operation with projective measurement. The advantage, however, is an absence of time-dependence in the procedure, with a passive steady-state monitoring of the error syndrome.

Note that the considered four-qubit Bacon-Shor code is a quantum error detecting code, while the smallest Bacon-Shor code for error correction contains nine qubits (not considered here). Therefore, so far our results are valid only for quantum error detection. While we guess that the QEC results for the nine-qubit (and higher) Bacon-Shor codes with continuous measurement are similar to the results presented here, this will require a separate analysis.

An operation of a usual quantum error detecting or correcting code (we consider only quantum memory for one logical qubit) assumes encoding a logical qubit into several physical qubits, keeping it for a relatively long time in the presence of decoherence, and then decoding it back into a logical qubit. (We do not consider fault-tolerant schemes in which logical operations are applied without decoding.) For simplicity, encoding and decoding are assumed to be perfect, so that we can focus on storage of quantum information only. In QEC the decoded logical qubit should always be ``handed back''; however, an error detecting code has also an option of not returning the logical qubit: the procedure is terminated  when an error is detected, since it cannot be corrected. Therefore, while the main performance characteristic for a QEC code is the probability of a logical error or the corresponding
logical error rate, a quantum error detecting code is characterized by two main parameters. The first parameter is the success probability (probability that the procedure is not terminated) or the corresponding success probability decay rate \cite{Knill-2005} (the rate of detected errors). For brevity we will call this rate the termination rate. The second parameter for a quantum error detecting code is the logical error probability (or the corresponding rate) conditioned on the absence of detected errors \cite{Knill-2005}. We will use the terminology of logical error rate, often omitting the word ``conditional''.
The termination rate is usually larger than the rate of errors in physical qubits (because the code is supposed to terminate operation when an error occurs), and it can be significantly larger due to ``false alarms'', when an error is indicated even though it actually did not occur. In this paper we calculate the logical error rate and the termination rate for the four-qubit Bacon-Shor code with continuous measurement and compare them with those for the conventional code operation with projective  measurements.

The paper in organized in the following way. In Sec. II we consider the conventional four-qubit Bacon-Shor code operated with projective measurements. We start with discussing the protocol (Sec.\ \ref{sec:system}) and its operation without errors (Sec.\ \ref{sec:proj-no-errors}), then discuss classification of single-qubit and two-qubit errors (Sec.\ \ref{sec:proj--error-class}), and then calculate the logical error rates and the termination rate for several models of decoherence (Sec.\ \ref{sec:proj-term-and-logical}). The four-qubit Bacon-Shor code with continuous measurements is analyzed in Sec.\ III. We start with an overview of the mathematical approach and results (Sec.\ \ref{sec:overview}). Then in more detail we analyze the evolution due to continuous measurement for a general state (Sec.\ \ref{sec:cont-gen-evolution}), without errors (Sec.\ \ref{sec:cont-no-error-evol}), and within error subspaces (\ref{sec:cont-error-evol}). The mapping due to single-qubit errors is discussed in Sec.\ \ref{sec:cont-error-mapping}, followed by calculation of logical error rates in Sec.\ \ref{sec:cont-logical-errors}. The false alarm rate and response time are analyzed in Sec.\ \ref{sec:cont-false-alarm}. Numerical results of Monte Carlo simulations are presented in Sec.\ \ref{sec:numerics}. Comparison between the operations with projective and continuous measurements is discussed in Sec.\ \ref{sec:cont-comparison}. Section \ref{sec:Conclusion} is the conclusion.

\section{Four-qubit Bacon-Shor code with projective measurements}\label{sec:codespace}

\subsection{System, protocol, and code space}\label{sec:system}

\begin{figure}[tb]
\includegraphics[width=7cm]{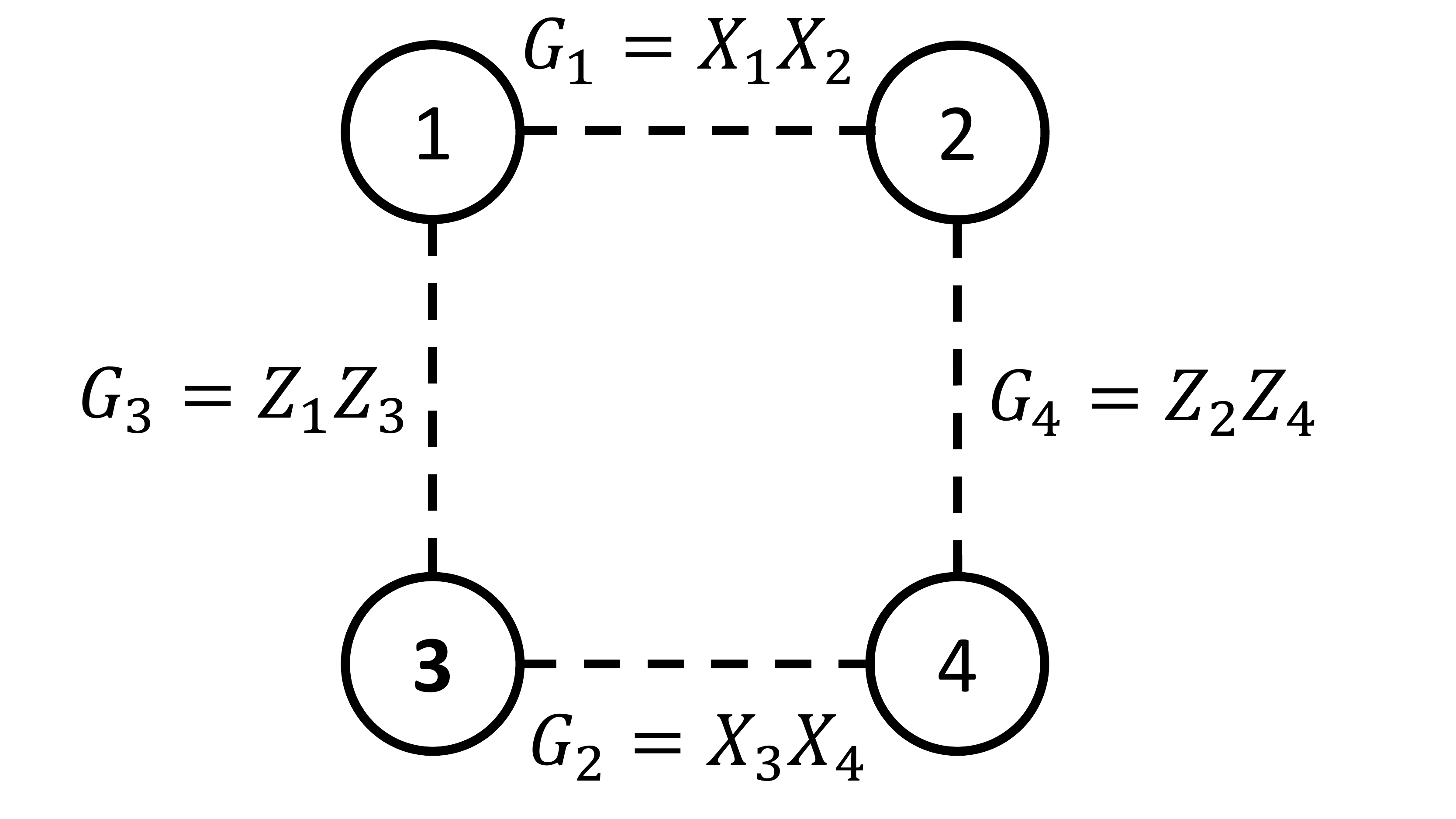}
\caption{The four-qubit Bacon-Shor code contains four physical qubits (shown by circles) and is based on measurement of four operators (dashed lines): $X_1X_2$, $X_3X_4$, $Z_1Z_3$, and $Z_2Z_4$, called gauge operators. In the conventional code operation they are measured in two steps, thus separating non-commuting pairs (Fig.\ 2), while in this paper we also analyze the case when all four operators are measured at the same time continuously.
}
\label{fig:4q}
\end{figure}

The four-qubit Bacon-Shor code  contains four qubits, labeled 1--4 in Fig.\ 1, and its conventional operation is based on projective  measurement of four two-qubit operators (gauge generators \cite{Poulin-2005,Bacon-2006}), for which we will interchangeably use the following notations:
    \be
    \begin{aligned}
    & X_1X_2 =X_{12}=G_1, \,\,\, &&  X_3X_4=X_{34}=G_2  ,
    \\
    & Z_1 Z_3 =Z_{13}=G_3, \,\,\, && Z_2 Z_4 = Z_{24}=G_4 ,
    \end{aligned}
   \label{4-operators}\ee
where $X=\sigma_x$ and $Z=\sigma_z$ are the Pauli matrices (similarly $Y=\sigma_y$) and indices of the Pauli operators indicate qubit numbering. Since the operators in the first and second lines of Eq.\ (\ref{4-operators}) do not commute with each other, they are measured sequentially. We will consider the version of the protocol shown in Fig.\ 2, in which each cycle of the protocol consists of two timesteps. The first time step of duration $\Delta t$ ends with instantaneous projective measurement of the operators $Z_{13}$ and $Z_{24}$, and the second timestep of the same duration $\Delta t$ ends with measurement of operators $X_{12}$ and $X_{34}$. (In principle,  duration of one of the timesteps can be almost zero, but we use $\Delta t$ for both of them, as more realistic for an experiment.) Note that $[X_{12},X_{34}]=[Z_{13},Z_{24}]=0$ and $\{X_{12},Z_{13}\}=\{X_{12},Z_{24}\}=\{X_{34},Z_{13}\}=\{X_{34},Z_{24}\}=0$, where $[\cdot ,\cdot]$ denotes commutator and $\{\cdot ,\cdot \}$ denotes anticommutator (these anticommutators are zero because the pairs contain exactly one common qubit). Also note that each of the four Pauli operators (\ref{4-operators}) has eigenvalues $\pm 1$, which correspond to the measurement results.

\begin{figure}[tb]
\includegraphics[width=70mm]{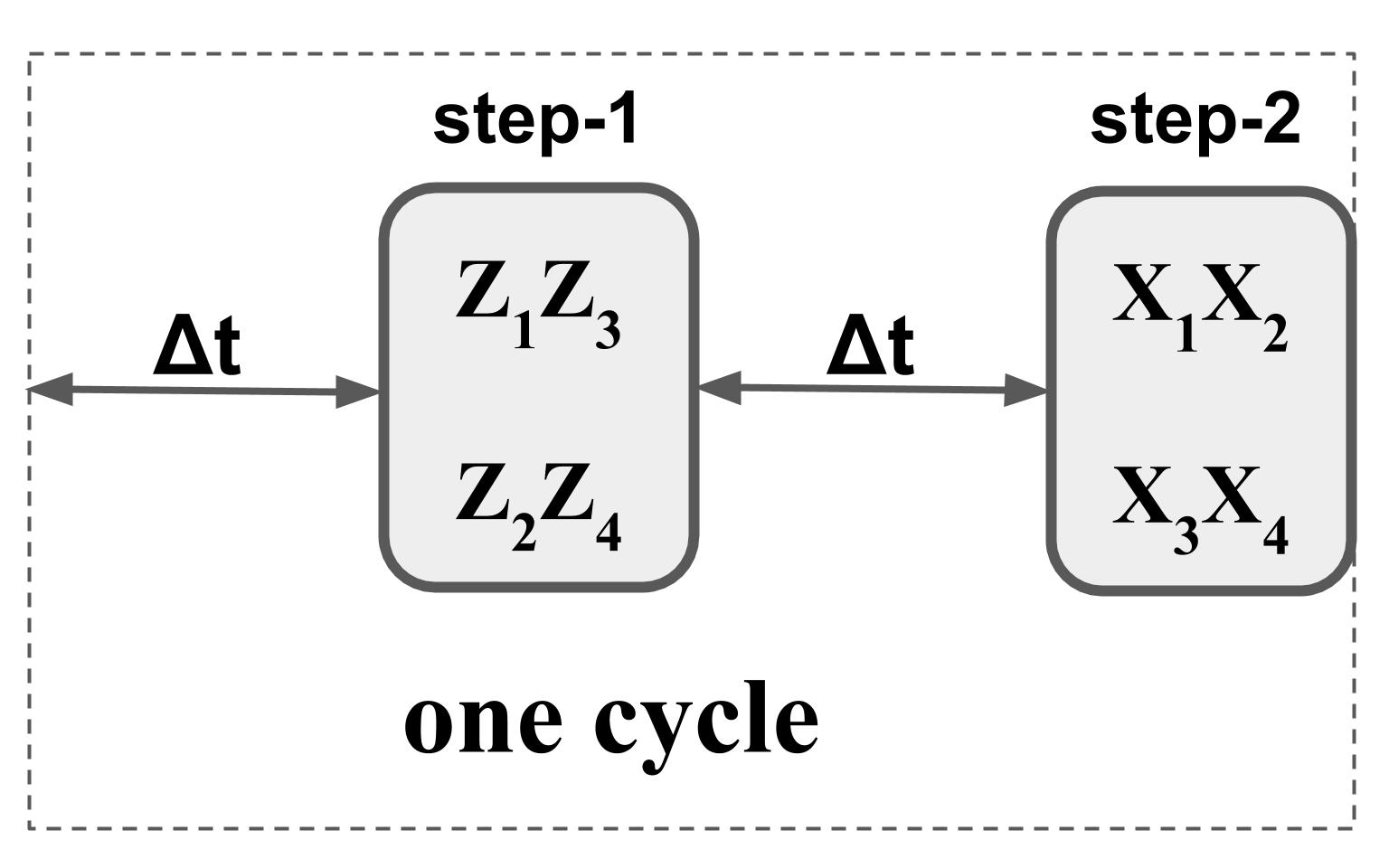}
\caption{One cycle of the code operation with projective measurement of gauge operators: at step 1 the operators $Z_1Z_3$ and $Z_2Z_4$ are measured instantaneously, and at step 2 the operators $X_1X_2$ and $X_3X_4$ are measured (also instantaneously). The duration of each timestep is $\Delta t$, so the cycle duration is $2\Delta t$. In the absence of errors, product of two measurement results ($\pm 1$) at each step is $+1$.}
\label{fig:2steps}
\end{figure}

The group generated by the measured gauge operators $G_k$ [Eq.\ (\ref{4-operators}), $k=1,2,3,4$] has an Abelian subgroup (stabilizer), whose every element also commutes with every $G_k$,  with generators
  \be
\begin{aligned}
& X_{\rm all} =X_1X_2X_3X_4=X_{12}X_{34},
\\
& Z_{\rm all} =Z_1Z_2Z_3Z_4=Z_{13}Z_{24}.
\end{aligned}
\label{stabilizers}\ee
These operators commute with all gauge operators $G_k$ and
have eigenvalues $\pm 1$, corresponding to parities of measurement results of $X$-type and $Z$-type operators in Eq.\ (\ref{4-operators}). The step-1 measurements (Fig.\ 2) project a 16-dimensional four-qubit state onto a 4-dimensional subspace belonging to one of the two 8-dimensional eigenspaces of $Z_{\rm all}$, while the step-2 measurements similarly  collapse the state into eigenspaces of $X_{\rm all}$. Since $[G_k,X_{\rm all}]=[G_k,Z_{\rm all}]=[X_{\rm all}, Z_{\rm all}]=0$ and measurement of an operator $G_k$ collapses four-qubit state with the projection operator $(\openone \pm G_k)/2$ ($\pm$ corresponds to the measurement result), an eigenstate of operators $X_{\rm all}$ and $Z_{\rm all}$ remains an eigenstate with the same eigenvalue after $G_k$ measurement (even though it changes the state). Therefore, after one cycle of the procedure (Fig.\ 2), any initial state is collapsed into one of four eigenspaces of operators $X_{\rm all}$ and $Z_{\rm all}$, and then remains in this eigenspace forever in the absence of decoherence. As already mentioned, operators $X_{\rm all}$ and $Z_{\rm all}$ are called the stabilizer generators of the code, while the measured operators $G_k$ are called gauge generators \cite{Poulin-2005,Bacon-2006}.

Different eigenvalues of operators $X_{\rm all}$ and $Z_{\rm all}$ divide 16-dimensional Hilbert space of four qubits into four orthogonal 4-dimensional subspaces. As usual, we choose the code space ${\cal Q}_0$ (``good'' subspace) as the eigenspace with eigenvalues $X_{\rm all}=+1$ and $Z_{\rm all}=+1$ (we use a somewhat sloppy notation for eigenvalues). For any state in the code space, the product of outcomes of $X_{12}$ and $X_{34}$ measurements is $+1$, and the product of $Z_{13}$ and $Z_{24}$ outcomes is also $+1$. The subspace with eigenvalues $X_{\rm all}=-1$ and $Z_{\rm all}=+1$ is denoted as ${\cal Q}_Z$ (this notation refers to the $Z$-error in a physical qubit, as discussed below). For any state in ${\cal Q}_Z$, the product of outcomes of $X_{12}$ and $X_{34}$ measurements  is $-1$, while for $Z_{13}$ and $Z_{24}$ the product is still $+1$.
Similarly, ${\cal Q}_X$ denotes the subspace with eigenvalues $X_{\rm all}=+1$ and $Z_{\rm  all}=-1$,  and subspace ${\cal Q}_Y$ has eigenvalues $X_{\rm all}=-1$ and $Z_{\rm all}=-1$.
The subspaces ${\cal Q}_X$, ${\cal Q}_Y$, and ${\cal Q}_Z$ are the ``error'' subspaces; the product of $-1$ for measurement outcomes at any step indicates an error.

Let us introduce the following orthonormal basis for the 4-dimensional code space ${\cal Q}_0$:
\begin{align}
\label{phi-1}
|\phi_1\rangle&=\left(|0000\rangle+|1111\rangle\right) /\sqrt{2},
\\\label{phi-2}
|\phi_2\rangle&=\left(|1100\rangle+|0011\rangle\right) /\sqrt{2},
\\\label{phi-3}
|\phi_3\rangle&=\left(|1010\rangle+|0101\rangle\right)/\sqrt{2} ,
\\\label{phi-4}
|\phi_4\rangle&=\left(|0110\rangle+|1001\rangle\right)/\sqrt{2} .
\end{align}
It is easy to see that $Z_{\rm all}=+1$ for all these vectors since the number of ones (and zeros) in each component is even. To check that $X_{\rm all}=+1$, we see that for each $|\phi_j\rangle$ ($j=1$--4) the two components in the superposition are complementary to each other, and the relative sign between the components is positive. The subspace ${\cal Q}_Z$ is spanned by the basis $\{Z_1 |\phi_j\rangle\}$  (equivalently, $Z_2$, $Z_3$ or $Z_4$ could be used, but we use $Z_1$). Similarly,  the subspace ${\cal Q}_X$ is spanned by $\{X_1 |\phi_j\rangle\}$ and ${\cal Q}_Y$ is spanned by $\{Y_1 |\phi_j\rangle\}$.

Initial state (encoded logical qubit) is always in the subspace ${\cal Q}_0$, and without decoherence it would remain in ${\cal Q}_0$ forever, so that the measurement outcomes at each step are either ``$++$'' or ``$--$'', with the product of $+1$ always. The outcomes ``$+-$'' or ``$-+$'' (with the product of $-1$) indicate an error. Since the four-qubit Bacon-Shor code is only an error detecting code, it cannot correct the error, and the procedure is terminated immediately after the product of $-1$ is obtained. Only in the case when the product of $+1$ was obtained for all measurement steps during $M\gg 1$ cycles (total duration of $T=2M\Delta t$), the quantum information is considered as preserved and decoded back into a logical qubit to be ``handed back''.

\subsection{Operation without errors}\label{sec:proj-no-errors}

The code space ${\cal Q}_0$ is 4-dimensional, but it is used to encode only one (2-dimensional) logical qubit. This is the usual feature of the subsystem codes \cite{Terhal-2015}. It is easy to see why additional dimensionality is needed in our case. After step-1 measurement, the four-qubit state can either contain a superposition of basis vectors $|\phi_1\rangle$ and $|\phi_3\rangle$ from Eqs.\ (\ref{phi-1})--(\ref{phi-4}) (if  the measurement outcome is ``$++$'') or a superposition of $|\phi_2\rangle$ and $|\phi_4\rangle$ (if  the measurement outcome is ``$--$''), but not a superposition of all four basis vectors. Moreover, if the outcome ``$++$'' is obtained, then one cycle later (with non-commuting step-2 measurement in between) the outcome will be ``$++$'' or ``$--$'' with equal probabilities. This additional degree of freedom (gauge) consumes two additional dimensions.

Let us encode the logical qubit
    \be
    |\psi\rangle_{\rm L} = \alpha\,|0\rangle_{\rm L} +\beta\, |1\rangle_{\rm L} =|\alpha,\beta\rangle_{\rm L}
    \label{logic-qubit}\ee
so that after step-1 measurement ($Z_{13}$ and $Z_{24}$), one of the following entangled four-qubit states can be obtained:
\begin{align}
\label{z+}
|z+\rangle&=\alpha\, |\phi_1\rangle+\beta\, |\phi_3\rangle ,
\\\label{z-}
|z-\rangle&=\alpha\, |\phi_2\rangle+\beta \, |\phi_4\rangle ,
\end{align}
where $|z+\rangle$ corresponds to the outcome ``$++$'' and $|z-\rangle$ corresponds to ``$--$''.
Then after step-2 measurement ($X_{13}$ and $X_{24}$) the two possible collapsed states are
\begin{align}
\label{x+}
|x+\rangle& =
\alpha\, \frac{|\phi_1\rangle+|\phi_2\rangle}{\sqrt{2}}
+\beta\, \frac{|\phi_3\rangle+|\phi_4\rangle}{\sqrt{2}} ,
\\\label{x-}
|x-\rangle& =
\alpha\, \frac{|\phi_1\rangle-|\phi_2\rangle}{\sqrt{2}}
+\beta\, \frac{|\phi_3\rangle-|\phi_4\rangle}{\sqrt{2}},
\end{align}
with $|x+\rangle$ corresponding to the outcome ``$++$'' and $|x-\rangle$ corresponding to ``$--$''. Then after step-1 measurement the produced state is again given either by Eq.\ (\ref{z+}) or Eq.\ (\ref{z-}), and the cycle repeats forever (assuming the absence of decoherence).

\begin{figure}[tb]
\includegraphics[width=5cm]{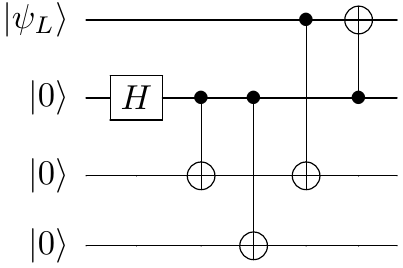}
\caption{Encoding circuit, which produces the state $|z+\rangle$ from the state $|\psi\rangle_{\rm L}|0\rangle | 0\rangle |0\rangle$. Decoding can be done by running the same circuit backwards after step-1 measurement with the outcome ``$++$''.  }
\label{fig:circuit}
\end{figure}

An encoding operation can be realized using a unitary, transforming the four-qubit state $|\psi\rangle_{\rm L}|0\rangle|0\rangle|0\rangle$ into $|z+\rangle$. For example, this can be done with the encoding unitary (Fig.\ \ref{fig:circuit})
    \begin{eqnarray}
&& U_{\rm enc}= \CNOT_{21} \CNOT_{13} \CNOT_{24} \CNOT_{23}H_2, \qquad
    \label{U-enc}\\
    && |z+\rangle = U_{\rm enc} \, (|\psi\rangle_{\rm L}|0\rangle|0\rangle|0\rangle),
    \end{eqnarray}
where indices are the qubit numbers, for $\CNOT_{ij}$ the first index is the control (the second is the target), and $H$ denotes Hadamard.
Eventual decoding can be done, for example, by applying the reversed unitary transformation $U_{\rm enc}^\dagger$ after step-1 measurement with the outcome ``$++$'', while for the outcome ``$--$'' we at first additionally apply operation $X_1X_2$, which transforms $|z-\rangle$ into $|z+\rangle$.

Note that the states $|z+\rangle$ and $|z-\rangle$ are orthogonal to each other (as well as the states $|x+\rangle$ and $|x-\rangle$) and
    \be
|x\pm\rangle = \frac{|z+\rangle \pm |z-\rangle}{\sqrt{2}}, \,\,\,\,\,
       |z\pm\rangle = \frac{|x+\rangle \pm |x-\rangle}{\sqrt{2}}.
    \label{x-pm-z-pm}\ee
Therefore all these four states belong to a 2-dimensional subspace (so-called gauge qubit), and the 4-dimensional subspace ${\cal Q}_0$ consists of such 2-dimensional subspaces, which are different for different logical qubit states (orthogonal to each other if logical qubit states are orthogonal).

\subsection{Error classification}\label{sec:proj--error-class}

While actual physical source of errors is a gradual decoherence, it is possible to think about it in terms of (possibly correlated) discrete $X$, $Y$, and $Z$ errors randomly applied to physical qubits \cite{N-C-book,Preskill-notes}. For simplicity we assume no correlation, and we also assume sufficiently small error rate, so that single-qubit errors are dominating, the two-qubit errors are next in the hierarchy, and so on.  A two-qubit error is realized when two independent errors occur in different qubits within the same timestep $\Delta t$ (or sometimes within $2\Delta t$); the rate of these errors will be discussed later, while in this section we essentially assume two errors occurring at the same time. We do not consider three-qubit errors because they are much less probable than two-qubit errors, and the important characteristics of the code are mainly determined by single-qubit and two-qubit errors. Note that we consider errors in the quantum channel setting, which assumes that projective measurements are ideal \cite{Bennett-1996}.

\subsubsection{Single-qubit errors}\label{sec:1-qubit-errors}

There are 12 possible single-qubit errors: $X_i$, $Y_i$, and $Z_i$, with $i=$1--4 labeling physical qubits. All these 12 types of errors are detectable by the code. It is easy to see that operator $X_i$ (with any $i$) applied to a state within the subspace ${\cal Q}_0$, moves it to the subspace ${\cal Q}_X$, since this flips the sign of the eigenvalue of $Z_{\rm all}$ (because $\{X_i,Z_i\}=0$ and therefore $\{X_i,Z_{\rm all}\}=0$) and does not affect the eigenvalue of $X_{\rm all}$ (because $[X_i,X_{\rm all}]=0$). Similarly, the error $Z_i$ (with any $i$) moves a state from ${\cal Q}_0$ to ${\cal Q}_Z$, since this flips $X_{\rm all}$ and does not change $Z_{\rm all}$. The errors $Y_i$ move a state from ${\cal Q}_0$ to ${\cal Q}_Y$ by flipping both $X_{\rm all}$ and $Z_{\rm all}$.

Therefore, all single-qubit errors are detectable with the following error syndromes:
\begin{itemize}
\item $X_i$ errors produce negative parity of outcomes (``$+-$'' or ``$-+$'') at step-1 measurements ($Z_{13}$ and $Z_{24}$), while producing usual positive parity  (``$++$'' or ``$--$'') at step-2 measurements ($X_{12}$ and $X_{34}$);
\item $Z_i$ errors produce positive parity at step-1 measurements and negative parity at step-2 measurements;
\item $Y_i$ errors produce negative parities for both step-1 and step-2 measurements.
\end{itemize}
Recall that without errors both parities are positive.

Since there are only three different error syndromes and 12 possible errors, the errors are not correctable, and the procedure should terminate when at least one parity of measured outcomes is negative. Therefore, the termination rate (success probability decay rate) for this code is the sum of rates for all single-qubit errors (these errors are dominating, so we do not need to include two-qubit and higher-order errors into the termination rate).

From the point of view of the termination rate, the considered quantum error detecting code is optimal (the termination rate is approximately equal to actual error rate). This is in contrast, for example, to the quantum error detection procedure based on uncollapsing \cite{Keane-2010,Kim-2012,Zhong-2014}, which experimentally demonstrated an increase of the qubit lifetime by a factor of three \cite{Zhong-2014}, but with a significantly smaller success rate than dictated by actual errors. As will be discussed in Sec.\ III, the four-qubit Bacon-Shor code with continuous measurements may have a significant contribution to the termination rate from ``false alarms''; then the code becomes non-optimal in this sense.

Note that any error operator $X_i$, $Y_i$ or $Z_i$ applied to a state in the subspace ${\cal Q}_0$, moves it to one of the error subspaces (${\cal Q}_X$, ${\cal Q}_Y$ or ${\cal Q}_Z$) and not to a superposition of states from different subspaces. If another error operator is applied after that, it also moves the state to one of the subspaces. Therefore, for any sequence of single-qubit error operators we never have a superposition of states from different subspaces.

\subsubsection{Two-qubit errors}\label{sec:2-qubit-errors}

There are $ (4\times 3/2) \times 3^2 = 54$ two-qubit error combinations,
which can be classified in the following way.
\begin{align}
\text{Harmless:} \,\,\,\,\,
X_1X_2,\,\, X_3X_4, \,\, Z_1Z_3,\,\, Z_2Z_4.
\label{log-harmless}\end{align}
When these operators are applied to the legitimate states (\ref{z+})--(\ref{x-}), the state either does not change (up to an overall phase) or changes within the gauge qubit subspace. Therefore, the effect is essentially unnoticeable and fully disappears after the next measurement. Note that the harmless combinations are the measured operators (\ref{4-operators}).

   \be \label{log-X}
\begin{aligned}
\text{Logical $X$ error:}\,\,\,\,&
X_1X_3,\,\, X_1X_4,\,\, X_2X_3,\,\, X_2X_4,
\\  & ~Y_1Y_3,~Y_2Y_4.
\end{aligned}
    \ee
For these combinations the state remains in the code space ${\cal Q}_0$ (and therefore, no error syndrome is produced); however, the logical qubit $|\alpha,\beta\rangle_{\rm L}$ transforms into $\pm|\beta,\alpha\rangle_{\rm L}$ (the overall phase $\pm$ is not important). It is easy to see that the state remains in ${\cal Q}_0$ because the operators (\ref{log-X}) commute with $X_{\rm all}$ and $Z_{\rm all}$ (since Pauli operators either commute or anticommute with each other). The transformation $|\alpha,\beta\rangle_{\rm L} \rightarrow\pm|\beta,\alpha\rangle_{\rm L}$ can be checked explicitly for the states $|z\pm\rangle$ and $|x\pm\rangle$ [Eqs.\ (\ref{z+})--(\ref{x-})] by using the following mapping: $|\phi_1\rangle \leftrightarrow |\phi_3\rangle$, $|\phi_2\rangle \leftrightarrow |\phi_4\rangle$ for $X_1X_3$ and $X_2X_4$;
$|\phi_1\rangle \leftrightarrow |\phi_4\rangle$, $|\phi_2\rangle \leftrightarrow |\phi_3\rangle$ for $X_1X_4$ and $X_2X_3$; $|\phi_1\rangle \leftrightarrow -|\phi_3\rangle$, $|\phi_2\rangle \leftrightarrow |\phi_4\rangle$ for $Y_1Y_3$ and $Y_2Y_4$ (these mappings may also exchange states $|z+\rangle \leftrightarrow |z-\rangle$ and/or $|x+\rangle \leftrightarrow |x-\rangle$). Note that the complementary combinations of operators ($X_1X_3$ and $X_2X_4$, also $X_1X_4$ and $X_2X_3$, also $Y_1Y_3$ and $Y_2Y_4$) have exactly the same action within ${\cal Q}_0$, and therefore it is sufficient to check a property only for one of the two complementary combinations. The equivalence can be easily proven by recalling that any state within ${\cal Q}_0$ is an eigenstate of $X_{\rm all}$, $Z_{\rm all}$, and $Y_{\rm all}=Y_1Y_2Y_3Y_4$ with eigenvalues of $+1$; therefore complements to $X_{\rm all}$, $Z_{\rm all}$, and $Y_{\rm all}$ are equivalent.

\begin{align}\label{log-Y}
\text{Logical $Y$ error:} \,\,\,\,\,
Y_1Y_4,\,\, Y_2Y_3.
\end{align}
For these combinations the state remains in the code space ${\cal Q}_0$, but the logical qubit $|\alpha,\beta\rangle_{\rm L}$ transforms into $\pm|\beta,-\alpha\rangle_{\rm L}$. This can be shown in the same way as for the combinations (\ref{log-X}), using the mapping $|\phi_1\rangle \leftrightarrow -|\phi_4\rangle$, $|\phi_2\rangle \leftrightarrow |\phi_3\rangle$. Note that the combinations $Y_1Y_4$ and $Y_2Y_3$ are complementary to each other, so only one of them needs to be checked.

    \be \label{log-Z}
\begin{aligned}
\text{Logical $Z$ error:}\,\,\,\,&
Z_1Z_2,\,\, Z_3Z_4,\,\, Z_2Z_3,\,\, Z_1Z_4,
\\ & ~Y_1Y_2,\,\, Y_3Y_4.
\end{aligned}
    \ee
For these combinations the state remains in ${\cal Q}_0$, but the logical qubit $|\alpha,\beta\rangle_{\rm L}$ transforms into $\pm|\alpha,-\beta\rangle_{\rm L}$. This can be shown via the mapping $|\phi_{1,2}\rangle \leftrightarrow |\phi_{1,2}\rangle$, $|\phi_{3,4}\rangle \leftrightarrow -|\phi_{3,4}\rangle$ for $Z_1Z_2$ and $Z_3Z_4$,
 $|\phi_{1,4}\rangle \leftrightarrow |\phi_{1,4}\rangle$, $|\phi_{2,3}\rangle \leftrightarrow -|\phi_{2,3}\rangle$ for $Z_1Z_4$ and $Z_2Z_3$, and $|\phi_1\rangle \leftrightarrow -|\phi_2\rangle$, $|\phi_3\rangle \leftrightarrow |\phi_4\rangle$ for $Y_1Y_2$ and $Y_3Y_4$.

The remaining $54-(4+6+2+6)=36$ two-qubit errors involve different error types and map a state from ${\cal Q}_0$ into one of the error subspaces; therefore, these combinations are detectable:
\begin{align}
\text{Detectable:} \,\,\,\,\,
X_iY_j,\,\, X_iZ_j,\,\, Y_iZ_j, \,\, i\neq j.
\end{align}

Note that we do not consider two-error combinations for the same qubit because they are equivalent to single-qubit operators and therefore are either harmless ($X_iX_i$, $Y_iY_i$, $Z_iZ_i$) or detectable ($X_iY_i$, $X_iZ_i$, $Y_iZ_i$).

\subsection{Termination and logical error rates}\label{sec:proj-term-and-logical}

In this section we consider several models of decoherence and calculate the termination and the logical error rates for the four-qubit Bacon-Shor code with projective measurements.

\subsubsection{Uncorrelated Markovian errors} \label{sec:Markovian-proj}

Let us consider first the usual model of errors, which assumes random Markovian errors of $X$, $Y$, and $Z$ types in each qubit, without correlations between the qubits. The rates of these 12 single-qubit errors may be all different and are denoted as $\Gamma_i^{(X)}$, $\Gamma_{i}^{(Y)}$, and $\Gamma_i^{(Z)}$, with index $i$ denoting the qubit.  We assume $\Gamma_i^{(X,Y,Z)}\Delta t \ll 1$, so that single-qubit errors are dominating, followed by two-qubit errors, and so on.

Note that physical decoherence produces mixed states, characterized by density matrices, while in this model an initially pure state remains pure, so it is sufficient to operate with wavefunctions for any given sequence of discrete errors. The averaging over these sequences, however, produces mixed states, corresponding to actual decoherence, so that the approach of discrete errors is essentially unraveling of physical decoherence. In general, if decoherence can be described as an evolution of the density matrix $\rho$ with the standard Lindblad form, involving single-qubit error operators $E_i$ ($E$ denotes a type of the process),
    \begin{eqnarray}
    && \dot{\rho} = \sum\nolimits_{i,E} \Gamma_i^{(E)} {\cal L}[E_i]\rho,  \,\,\,
        \label{Lindblad-1}\\
    && {\cal L}[A]\rho \equiv A\rho A^\dagger -\frac{1}{2} (A^\dagger A \rho +\rho A^\dagger A),
    \label{Lindblad-2}\end{eqnarray}
then it can be replaced (unravelled) with the following  ``jump/no-jump'' evolution (see, e.g., \cite{Haroche-book,Korotkov-2013}). The ``jumps'' with Kraus operators $E_i$  are randomly applied with the rates $\Gamma_i^{(E)} {\rm Tr} (E_i^\dagger E_i\rho)$ (the state is normalized after each jump), while the ``no-jump'' evolution (essentially the quantum Bayesian update) for a short time $\delta t$ is described by the Kraus operator $\openone - \sum_{i,E} \Gamma_{i}^{(E)} \delta t \, E_i^\dagger E_i/2$ (the state requires normalization after the no-jump evolution as well). For the model considered in this section, $E=X,$ $Y$, and $Z$, i.e., the error operators are of the Pauli-matrix type. In this case $E_i^\dagger E_i=\openone$; therefore the jump rates do not depend on the state and are equal to $\Gamma_i^{(E)}$, while the no-jump evolution is trivial, so that we do not need to consider it explicitly.

As discussed above, any single-qubit error is detected at the next or second-next measurement step and the procedure is terminated, unless another single-qubit error occurs before the detection and returns the state into subspace ${\cal Q}_0$ (leading to a non-detectable two-qubit error discussed above). Neglecting the non-detectable two-qubit (and higher-order) errors, the termination rate $\gamma_{\rm term}$ is the sum of all single-qubit error rates,
    \be
    \gamma_{\rm term} = \sum\nolimits_i \left[ \Gamma_i^{(X)}+\Gamma_{i}^{(Y)}+\Gamma_i^{(Z)} \right] .
    \label{Markovian-term}\ee
Note that the non-detectable two-qubit errors slightly decrease the termination rate, with the relative correction to Eq.\ (\ref{Markovian-term}) on the order of $\gamma_{\rm term} \Delta t \ll 1$. The probability that the procedure is not terminated until the end of the operation of duration $T$ is the ``success'' (``survival'') probability
    \be
    P_{\rm success} =\exp (-\gamma_{\rm term} T).
    \label{success-prob}\ee

If two errors occur sufficiently close in time and the second error returns the state back to the subspace ${\cal Q}_0$, then the procedure does not detect any error. However, as discussed in Sec.\ \ref{sec:2-qubit-errors}, it is possible that the state changes significantly, so that the logical qubit acquires $X$, $Y$ or $Z$ error. To find the rate of logical $X$-errors, we use the combinations in Eq.\ (\ref{log-X}) and notice that a combination $X_iX_j$ will not be detected if both errors occur within the same cycle $2\Delta t$ between neighboring step-1 measurements ($Z_{13}$ and $Z_{24}$). The corresponding rate is then $2\Delta t\,\Gamma_i\Gamma_j$. The $Y_iY_j$ combinations in Eq.\ (\ref{log-X}) will be undetected only if both errors occur within the half-cycle $\Delta t$ between the neighboring measurements. The corresponding rate is $\Delta t\, \Gamma_i\Gamma_j$. Summing over all scenarios, we obtain the rate of logical $X$-error,
    \begin{eqnarray}
    && \gamma_X= \Delta t \left[2(\Gamma_1^{(X)}+\Gamma_2^{(X)})(\Gamma_3^{(X)}+\Gamma_4^{(X)}) \right.
    \nonumber \\
    &&\hspace{1.6cm} \left. +\Gamma_{1}^{(Y)}\Gamma_{3}^{(Y)}+\Gamma_{2}^{(Y)}\Gamma_{4}^{(Y)} \right] .
    \label{proj-X-Mark}\end{eqnarray}
Similarly, for logical $Y$-error we use combinations in Eq.\ (\ref{log-Y}) and obtain the rate
    \be
       \gamma_Y= \Delta t \left[ \Gamma_{1}^{(Y)}\Gamma_{4}^{(Y)}+\Gamma_{2}^{(Y)}\Gamma_{3}^{(Y)} \right] .
    \label{proj-Y-Mark}\ee
For the rate of logical $Z$-error we use Eq.\ (\ref{log-Z}) and obtain
    \begin{eqnarray}
    && \gamma_Z= \Delta t \left[2(\Gamma_1^{(Z)}+\Gamma_3^{(Z)})(\Gamma_2^{(Z)}+\Gamma_4^{(Z)}) \right.
    \nonumber \\
    &&\hspace{1.6cm} \left. +\Gamma_{1}^{(Y)}\Gamma_{2}^{(Y)}+\Gamma_{3}^{(Y)}\Gamma_{4}^{(Y)} \right] .
    \label{proj-Z-Mark}\end{eqnarray}

If all single-qubit rates are equal, $\Gamma_i^{(X)}=\Gamma_i^{(Y)}=\Gamma_i^{(Z)}=\Gamma_{\rm d}/3$ (depolarizing channel \cite{Preskill-notes}), then
    \be
    \gamma_X=\gamma_Z = \frac{10}{9}\, \Gamma_{\rm d}^2\Delta t, \,\,\,
       \gamma_Y = \frac{2}{9}\, \Gamma_{\rm d}^2\Delta t.
    \ee

Let us introduce the total logical error rate (conditioned on no errors detected \cite{Knill-2005})
    \be
   \gamma_{\rm L} = \gamma_X+\gamma_Y+\gamma_Z.
    \label{gamma-L}\ee
For the depolarizing channel with $\Gamma_i^{(X)}= \Gamma_i^{(Y)}= \Gamma_i^{(Z)}= \Gamma_{\rm d}/3$, we have
   \be
    \gamma_{\rm L}=\frac{22}{9}\, \Gamma_{\rm d}^2 \, \Delta t,
    \label{gamma-L-depol-proj}\ee
which is much smaller than the error rate $\Gamma_{\rm d}$ without encoding if $\Gamma_{\rm d}\Delta t\ll 1$.

Note that the logical error rates are proportional to the time $\Delta t$ between the measurements (while $\gamma_{\rm term}$ does not depend on $\Delta t$). This is as it should be expected for a code with the logical errors caused by two-qubit errors. The operation of the code improves with smaller $\Delta t$, whose choice therefore should be based on technical (experimental) limitations.

\subsubsection{Pure dephasing}\label{sec:dephasing-proj}

The results of the previous section can be readily applied to analyze the effects of pure dephasing of physical qubits. Let us denote the rate of pure dephasing of $i$th qubit as $\Gamma_{\varphi,i}$ and assume no other sources of decoherence.
Effect of pure dephasing is equivalent to random $Z$-jumps with the rate $\Gamma_i^{(Z)}=\Gamma_{\varphi,i}/2$ (e.g., \cite{Korotkov-2013}). Therefore, we can use Eqs.\ (\ref{Markovian-term})--(\ref{proj-Z-Mark}) to obtain the termination and logical error rates,
    \begin{eqnarray}
    && \gamma_{\rm term} = \sum\nolimits_i \Gamma_{\varphi,i}/2,
    \label{term-proj-deph}\\
    && \gamma_X=0, \,\,\, \gamma_{Y}=0, \,\,\,
     \label{proj-XY-deph}\\
     && \gamma_{Z}= \Delta t \, (\Gamma_{\varphi,1}+\Gamma_{\varphi,3})(\Gamma_{\varphi,2}
     +\Gamma_{\varphi,4})/2.
    \label{proj-Z-deph}\end{eqnarray}

In particular, in the case of equal dephasing in all four qubits, $\Gamma_{\varphi,i}=\Gamma_\varphi$, we obtain the total logical error rate (with no detected errors)
    \be
\gamma_{\rm L}=\gamma_Z=2\,\Gamma_\varphi^2\,\Delta t,
    \label{gamma-L-proj-deph}\ee
which can be compared with the logical error rate
$\Gamma_\varphi /2$ without encoding.

\subsubsection{Energy relaxation}\label{sec:relax-proj}

Now let us discuss the model of zero-temperature energy relaxation (amplitude damping), $|1\rangle\rightarrow |0\rangle$, relevant to superconducting qubits. We assume uncorrelated energy relaxation of $i$th qubit with the rate $\mu_i\equiv 1/T_{1,i}\ll (\Delta t)^{-1}$.

This decoherence can be unraveled as ``jump/no-jump'' process (e.g., \cite{Haroche-book,Korotkov-2013}), consisting of random ``jumps'' caused by application of lowering Kraus operators $\sigma_{-,i}$ with the rates $\mu_{i}{\rm Tr} ( \sigma_{+,i}\sigma_{-,i}\rho)$ and ``no-jump'' evolution with Kraus operator $\openone-\sum_i (\mu_{i}\,\delta t/2)\, \sigma_{+,i}\sigma_{-,i}$ for an infinitesimal duration $\delta t$ with no jumps.
Here $\sigma_{-,j}=\sigma_{+,j}^\dagger=(X_j+\imath Y_j)/2$ (this definition assumes the state $|0\rangle$ to be at the top and $|1\rangle$ at the bottom of a spinor), and instead of the four-qubit density matrix $\rho$, we can  think in terms of a wavefunction. Note that $\sigma_{+,i}\sigma_{-,i}=(\openone -Z_i)/2$.

Let us start with jump processes (as discussed later, the no-jump processes do not affect the termination and logical error rates).
Since the states (\ref{z+})--(\ref{x-}) contain equal superpositions of $|0\rangle$ and $|1\rangle$ for each qubit, the ``jump'' rate in $i$th qubit is $\mu_i/2$. A single-qubit jump is necessarily detected since the resulting state is a superposition of states in subspaces ${\cal Q}_X$ and ${\cal Q}_Y$. Therefore, the termination rate is
    \be
    \gamma_{\rm term} = \sum\nolimits_i \mu_i/2,
    \label{term-proj-rel}\ee
independently of the logical state.

A logical error may occur when two jumps in different qubits occur within the same half-cycle $\Delta t$ or in the neighboring half-cycles. If the jumps in qubits $i$ and $j$ occur within the same $\Delta t$, then a legitimate wavefunction (\ref{z+})--(\ref{x-}) is multiplied by $(X_i+\imath Y_i)(X_j+\imath Y_j)/4$ (squared norm is proportional to probability). Since the combinations $X_iY_j$ are detectable, we are left with $(X_iX_j-Y_iY_j)/4$, which lead to logical errors, as discussed in Sec.\ \ref{sec:2-qubit-errors}. We need to be careful in applying Eqs.\ (\ref{log-harmless})--(\ref{log-Z}) to these combinations because of superposition of states produced by $(X_iX_j-Y_iY_j)/4$ and therefore possible interference effects. However, for most of the qubit pairs there is no interference because $X_iX_j$ and $Y_iY_j$ produce states in different subspaces corresponding to different logical states [see Eqs.\ (\ref{log-harmless})--(\ref{log-Z})]. Only for the qubit pair 1 and 3 (and complementary pair 2 and 4) the states may interfere: combinations $X_1X_3$ and $Y_1Y_3$ both produce logical $X$-error. By applying $(X_1X_3-Y_1Y_3)/4$ to the states $|z\pm\rangle$, we find transformations $|z+\rangle \rightarrow (2/4)|z+\rangle_{\alpha\leftrightarrow \beta}$, $|z-\rangle \rightarrow 0$, so the interference occurs, but its effect disappears after averaging over states $|z\pm\rangle$. Similarly, this operator produces transformations $|x+\rangle\rightarrow (\sqrt{2}/4) |z+\rangle_{\alpha\leftrightarrow \beta}$, $|x-\rangle\rightarrow (\sqrt{2}/4) |z+\rangle_{\alpha\leftrightarrow \beta}$, which correspond to the same probabilities, as without interference. Thus, interference between terms $X_iX_j$ and $Y_iY_j$ in producing logical errors is not important, and we can simply use Eqs.\ (\ref{log-harmless})--(\ref{log-Z}) to calculate probabilities of logical errors. For example, for qubits 1 and 2, the probability to have two jumps within $\Delta t$ is $(\mu_1\Delta t/2)(\mu_2\Delta t/2)$, and this produces logical $Z$-error with probability $1/4$ and no error (harmless combination) with probability $1/4$ (with probability $1/2$ the error will be detected). This produces a rate $\mu_1\mu_2\Delta t/16$ of logical $Z$-error. As another example, for qubits 1 and 3, the probability of two jumps within $\Delta t$ is $(\mu_1\Delta t/2)(\mu_3\Delta t/2)$, leading to logical $X$-error with probability $2/4$, thus producing the rate  $\mu_1\mu_3\Delta t/8$.
 Calculation of rates for other qubit pairs is similar.

If the jumps in qubits $i$ and $j$ occur in neighboring half-cycles $\Delta t$ separated by step-1 measurements ($Z_{13}$ and $Z_{24}$, then the error will necessarily be detected. However, if the half-cycles are separated by step-2 measurements ($X_{12}$ and $X_{34}$), then in the case of no error detected, the term $X_iX_j/4$ survives and may lead to logical error. We need to add these logical error rates to the rates due to both jumps occurring within the same $\Delta t$. Thus, we obtain the following rates of the logical errors,
    \begin{eqnarray}
      && \gamma_X= \frac{\Delta t}{16} (3\mu_1\mu_3 +2\mu_1\mu_4+3\mu_2\mu_4+2\mu_2\mu_3),\qquad
      \label{proj-X-rel} \\
       && \gamma_{Y}=  \frac{\Delta t}{16} (\mu_1\mu_4 +\mu_2\mu_3),
       \label{proj-Y-rel} \\
     && \gamma_{Z}= \frac{\Delta t}{16} (\mu_1\mu_2 +\mu_3\mu_4).
     \label{proj-Z-rel}
    \end{eqnarray}

Note that these rates coincide with the results (\ref{proj-X-Mark})--(\ref{proj-Z-Mark}) if we use $\Gamma_i^{(X)}=\Gamma_i^{(Y)}=\mu_i/4$ and $\Gamma_i^{(Z)}=0$. This similarity follows from unimportance of the discussed above interference between the effects of the terms $X_iX_j/4$ and $Y_iY_j/4$. However, the superposition $X_iX_j/4-Y_iY_j/4$ produces non-zero off-diagonal elements of the quantum process (tomography) matrix $\chi$ \cite{N-C-book,Kofman-2009} for the logical qubit. In particular, the relaxation jumps in qubits 1 and 2 (or 3 and 4) within the same $\Delta t$ contribute to both $Z$-error and harmless process, thus leading to the contribution $\chi_{IZ}/T=(1/16)(\mu_1\mu_2+\mu_3\mu_4)\Delta t$, where $T$ is the total duration of the process and we define $\chi$ as the conditional process matrix, selecting only the realizations with no detected errors. Other qubit pairs do not contribute to the off-diagonal elements of $\chi$ after averaging over states $|z\pm\rangle$ and $|x\pm\rangle$ [it is easier to analyze these contributions by using considered later approach of Eq.\ (\ref{error-correspondence}) and Fig.\ \ref{fig:continuous-scheme} and averaging over the gauge qubit state]. Note that by definition $\chi_{XX}/T=\gamma_X$, $\chi_{YY}/T=\gamma_Y$, and $\chi_{ZZ}/T=\gamma_Z$ (neglecting ``initial decoherence'' \cite{Falci-2005}).

Now let us discuss effect of the no-jump evolution, corresponding to the Kraus operator  $\openone -\delta t \sum_{i}\mu_i (\openone -Z_i)/4$ for infinitesimal $\delta t$. The product of these operators within the cycle time $2\Delta t$ between step-2 measurements ($X_{12}$ and $X_{34}$), produces the Kraus operator (to second order) $\openone -(\Delta t/2) \sum_{i}\mu_i (\openone -Z_i)+ (\Delta t/2)^2\sum_{i<j}\mu_i\mu_j (1-Z_i)(1-Z_j)+(\Delta t)^2/8\sum_i \mu_i^2 (1-Z_i)^2$. Hence, the step-2 measurements will detect $Z$-error with probability (in the leading order) $(\Delta t/2)^2(\sum_i\mu_i)^2$, which is proportional to $(\Delta t)^2$ and therefore can be neglected in calculation of the termination rate (\ref{term-proj-rel}). When no error is detected, the state self-corrects by eliminating terms $Z_i$ from the Kraus operator; however, the product-terms $Z_iZ_j$ are not eliminated, they accumulate for the whole duration $T$ of the process as $(T\Delta t/8)\sum_{i<j}\mu_i\mu_j Z_iZ_j$. This leads to logical $Z$-error [see Eq.\ (\ref{log-Z})] with probability $T^2 (\Delta t/8)^2 (\mu_1\mu_2+\mu_3\mu_4)^2$ [the terms with combinations $\mu_1\mu_4$ and $\mu_2\mu_3$ do not contribute because of averaging over the gauge qubit states, as can be understood by using Eq.\ (\ref{error-correspondence}) and Fig.\ \ref{fig:continuous-scheme}]. This is a ``coherent'' error \cite{Korotkov-2013}, which scales as $T^2$ with time and therefore cannot be characterized by a rate. However, it is easy to check that for a typical duration of the code operation, $T\sim \mu_i^{-1}$, this error is still much smaller than the errors accumulated with the rates (\ref{proj-X-rel})--(\ref{proj-Z-rel}). Therefore, the logical errors due to no-jump process can be neglected. We can also check that combinations of no-jump terms $Z_i$ with single-jump operators $(X_j+\imath Y_j)/2$ always produce detectable errors and therefore do not contribute to logical errors.

Even though logical errors due to no-jump evolution can be neglected, the coherent $Z$-error produces the contribution $\chi_{IZ}/T=(\Delta t/8) (\mu_1\mu_2+\mu_3\mu_4)$ to off-diagonal element of the logical quantum process matrix $\chi$. Combining it with the discussed above contribution from the double-jump processes, we obtain (assuming no detected errors)
    \be
    \chi_{IZ}=\chi_{ZI}=\frac{3}{16} (\mu_1\mu_2+\mu_3\mu_4) \, T \Delta t =3\chi_{ZZ}.
    \label{proj-IZ-rel}\ee

We have checked numerically Eqs.\ (\ref{proj-X-rel})--(\ref{proj-IZ-rel}) and found a very good agreement. In numerical calculations  we used Lindblad-form evolution of the 4-qubit density matrix due to energy relaxation of qubits and also used projectors onto the corresponding subspaces to simulate step-1 and step-2 measurements with results ``$++$'' or ``$--$''. The two projectors were added incoherently to take into account both measurement results; for a step-1 measurement we used projector $\rho_{\rm after}=\Pi_{++}^{G34}\rho_{\rm before}\Pi_{++}^{G34} +\Pi_{--}^{G34}\rho_{\rm before}\Pi_{--}^{G34}$, where $\Pi_{++}^{G34}=(\openone +G_3)(\openone+G_4)/4$, $\Pi_{--}^{G34}=(\openone -G_3)(\openone-G_4)/4$, while  $\rho_{\rm before}$ and $\rho_{\rm after}$ are the density matrices before and after a step-1 measurement. Similar procedure with projectors  $\Pi_{++}^{G12}=(\openone +G_1)(\openone+G_2)/4$ and $\Pi_{--}^{G12}=(\openone -G_1)(\openone-G_2)/4$ was used for a step-2 measurement. The process matrix $\chi$ as a function of the number of cycles was calculated using four initial logic states. We checked that the survival (success) probability is practically the same for all initial states [also checking the probability decay rate (\ref{term-proj-rel})] and then normalized the trace-non-preserving $\chi$ to find the process matrix conditioned on the absence of detected errors. We checked that the diagonal elements $\chi_{XX}$, $\chi_{YY}$, and $\chi_{ZZ}$ of the trace-preserving $\chi$ are given by the rates (\ref{proj-X-rel})--(\ref{proj-Z-rel}) multiplied by the total duration $T$ of the process, and the only non-zero off-diagonal element is given by Eq.\ (\ref{proj-IZ-rel}). In a similar way we numerically analyzed the case of pure dephasing; we checked Eqs.\ (\ref{term-proj-deph})--(\ref{proj-Z-deph}) and also checked the absence of non-zero off-diagonal elements of the process matrix $\chi$.

\section{Four-qubit Bacon-Shor code with continuous measurements}

Now let us consider the four-qubit Bacon-Shor code, in which the sequential projective measurement of four gauge operators $G_k$ (Fig.\ 2) is replaced with their simultaneous continuous measurement. We will first discuss the approach to this problem and results qualitatively, and then present the detailed analysis.

\subsection{Overview}\label{sec:overview}

Each measured gauge operator $G_k$ [Eq.\ (\ref{4-operators})] has eigenvalues $\pm 1$, which divide the 16-dimensional Hilbert space into two 8-dimensional subspaces, so that the measurement of $G_k$ distinguishes these two subspaces. In this sense continuous measurement of $G_k$ is similar to measurement of a qubit (two subspaces instead of two states), and we can use many previously obtained results for continuous measurement of a qubit. We will mainly use the quantum Bayesian formalism \cite{Korotkov-1999-2001,Korotkov-2003,Korotkov-2011}, which is essentially equivalent to the theory of quantum trajectories \cite{Wiseman-Milburn-1993,Carmichael-1993,Wiseman-Milburn-book}.

Using this formalism, we will show that in the absence of errors, continuous measurement of four gauge operators $G_k$ leads to the four-qubit state evolution
    \be
    |\psi (t)\rangle = a(t) \, |z+\rangle +b(t) \, |z-\rangle,
    \label{psi-a-b}\ee
where $a(t)$ and $b(t)$ are (in general complex) numbers with condition $|a|^2+|b|^2=1$, and states $|z\pm\rangle$  are given by Eqs.\  (\ref{z+}) and (\ref{z-}). Thus, continuous measurement of the gauge operators causes evolution of the ``gauge qubit'' $|a,b\rangle_{\it g}$, while not disturbing the logical qubit $|\alpha,\beta\rangle_{\rm L}$, which determines the basis states $|z\pm\rangle$. The typical timescale of the gauge qubit evolution is comparable to the collapse timescale $\tau_{\rm m}$ (so-called ``measurement time''\cite{Korotkov-2003}), corresponding to $G_k$ measurements ($\tau_{\rm m}^{-1}$ characterizes  measurement strength, and we assume equal strength for all four measurement channels).
Note that strictly speaking the result (\ref{psi-a-b}) is valid only when ideal (quantum-limited) detectors are used for $G_k$ measurement, while for non-ideal detectors the state should instead be described as an evolving density matrix of the gauge qubit. However, with a logical trick discussed later, it is still possible to use wavefunctions  to understand the code operation, while quantitative analysis can be done either using wavefunctions or density matrices.

In the 2-dimensional subspace spanned by $|z+\rangle$ and $|z-\rangle$, measurement of the operator $G_3=Z_{13}$ is simply $Z$-measurement of the effective (gauge) qubit, for which
    $|0\rangle_{\it g}=|z+\rangle$ and $|1\rangle_{\it g} =|z-\rangle$.
Similarly, measurement of $G_4=Z_{24}$ is also a $Z$-measurement of the same gauge qubit. In contrast, measurement of $G_1=X_{12}$ (or $G_2=X_{34}$) measures $X$-component of the gauge qubit (\ref{psi-a-b}). Thus, simultaneous continuous measurement of four operators $G_k$ is simply a continuous measurement of $Z$ and $X$ components of the gauge qubit (two $Z$-measurements and two $X$-measurements). The theory of such simultaneous $X$ and $Z$ measurement of a qubit has been developed in Ref.\ \cite{Ruskov-2010}, and it was experimentally realized in Ref.\ \cite{Hacohen-Courgy-2016}. In particular, in the absence of phase back-action from measurement and for equal strength of all measurements, the state (\ref{psi-a-b}) evolves as in the standard diffusion along the great circle of the Bloch sphere with real $a$ and $b$.

Note that sequential projective measurement of gauge operators $G_k$ leads to state jumps between $|z\pm\rangle$ and $|x\pm\rangle$ [see Eq.\ (\ref{x-pm-z-pm})], while continuous measurement of $G_k$  replaces jumps with continuous evolution; otherwise the state evolution in both cases is similar. However, an important difference is that in the projective case the measurement results are $\pm 1$, i.e., discrete, while continuous measurement of operators $G_k$ produces four noisy signals: $I_{X12}(t)$, $I_{X34}(t)$, $I_{Z13}(t)$, and $I_{Z24}(t)$ (here the subscripts indicate the measured operators). The positive parity of the projective measurement results, $X_{12}X_{34}=+1$ and $Z_{13}Z_{24}=+1$, in this case is replaced with positive {\it cross-correlators} for the noisy signals: $\langle I_{X12}(t) \, I_{X34}(t) \rangle=+1$ and $\langle I_{Z13} (t) \, I_{Z24}(t) \rangle=+1$. Thus, analysis of the Bacon-Shor code operation with continuous measurements significantly relies on the results for signal correlators in continuous qubit measurement \cite{Korotkov-2001-spectrum,Korotkov-2011-correlation,Atalaya-2016}.

A single-qubit error $X_i$ moves the four-qubit state from the subspace ${\cal Q}_0$ to the orthogonal subspace ${\cal Q}_X$ (see Sec.\ \ref{sec:system}). Any state in this subspace has negative $Z$-correlator, $\langle I_{Z13} (t) \, I_{Z24}(t) \rangle=-1$, while $X$-correlator is still positive, $\langle I_{X12}(t) \, I_{X34}(t) \rangle=+1$. Similarly, a $Z_i$ error moves the state from ${\cal Q}_0$ to ${\cal Q}_Z$, for which $\langle I_{X12}(t) \, I_{X34}(t) \rangle=-1$ and $\langle I_{Z13} (t) \, I_{Z24}(t) \rangle=+1$, and finally a $Y_i$ error moves to the subspace ${\cal Q}_Y$, in which both correlators are negative, $\langle I_{X12}(t) \, I_{X34}(t) \rangle=\langle I_{Z13} (t) \, I_{Z24}(t) \rangle=-1$. Thus, the cross-correlators for the output signals allow us to detect errors. Unfortunately, the products of noisy signals are very noisy, and therefore monitoring of a cross-correlator in real time is not easy. We will construct approximate cross-correlators $C_{12}(t)$ and $C_{34}(t)$ via double-integration in time of the corresponding pairs of the output signals (the indices here correspond to numbering of operators $G_k$). The integration kernel will be characterized by two time scales: parameter $\tau_{\rm c}$ for the integration over time difference in the two channels and much longer parameter $T_{\rm c}$ for integration over the mean time. The parameter $\tau_{\rm c}$ can be optimized, while the parameter $T_{\rm c}$ affects a trade-off between the noisiness of the approximate correlators ($C_{12}$ and $C_{34}$) and their average time of response, $T_{\rm R}$, to jumps of the actual correlators from $+1$ to $-1$. A short $T_{\rm c}$ makes $C_{12}$ and $C_{34}$  too noisy and therefore they will often cross zero or another threshold, erroneously indicating an error (false alarm). On the other hand, a long $T_{\rm c}$ makes $C_{12}$ and $C_{34}$ too slow, so that they report an error with a long delay $T_{\rm R}$ after it actually occurred, thus increasing the probability of logical error, as discussed below.

\begin{figure}[tb]
\includegraphics[width=8.5cm,trim=4cm 4.5cm 13cm 2.5cm,clip=true]{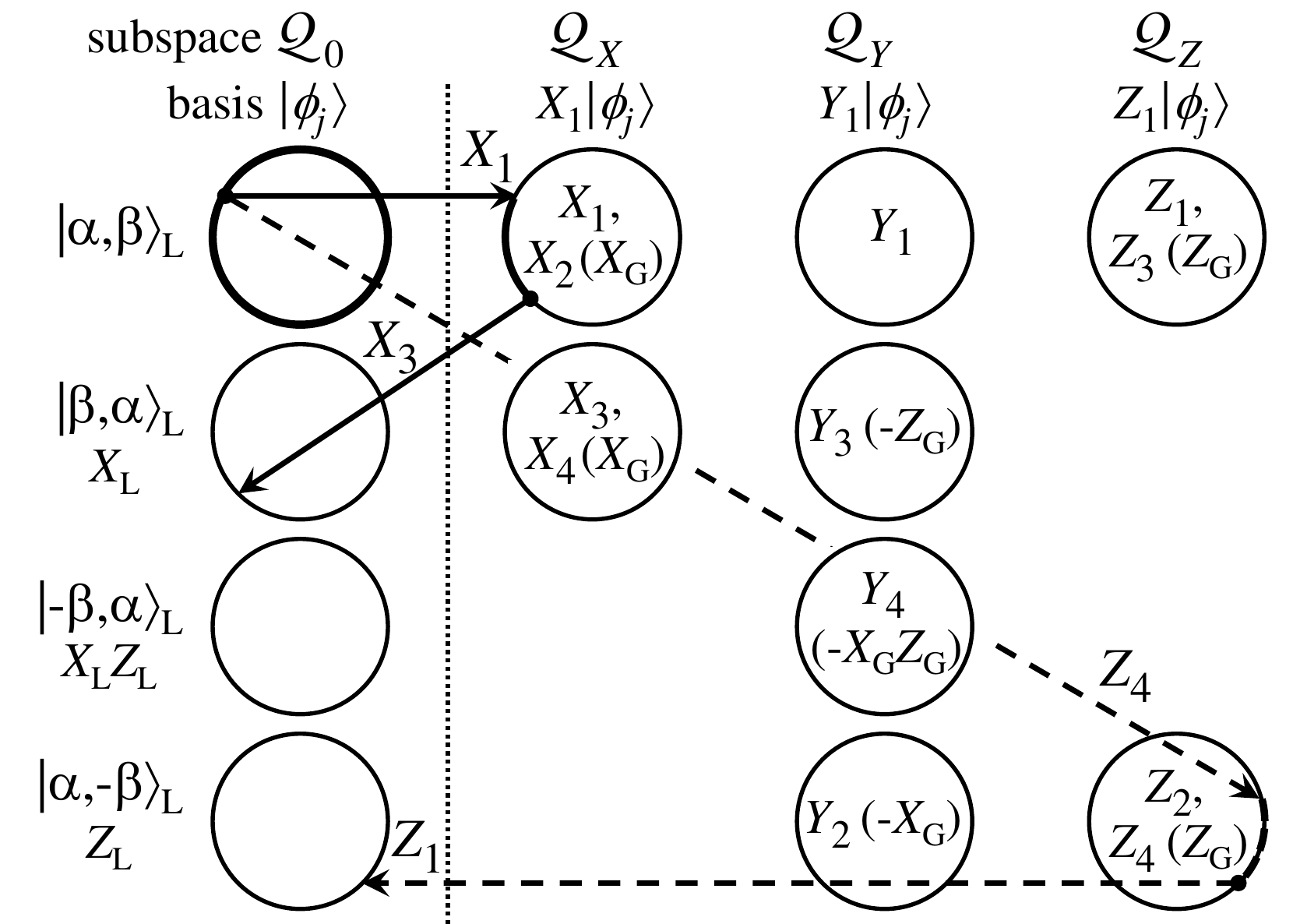}
\caption{State evolution and errors with continuous measurement. Thick-line circle illustrates diffusive evolution of the legitimate state [Eq.\ (\ref{psi-a-b})]  within 2-dimensional subspace of the gauge qubit due to measurement. The 12 single-qubit errors move the legitimate state to 8 other 2-dimensional subspaces in ${\cal Q}_X$, ${\cal Q}_Y$, and ${\cal Q}_Z$ with the mapping described by Eq.\ (\ref{error-correspondence}) -- see operators written inside the circles, the corresponding gauge qubit operations are in parentheses. If the second error occurs before the first error is detected (i.e., within the response time $T_{\rm R}$), it may move the state back to ${\cal Q}_0$, so that both errors remain undetected (dotted line separates ${\cal Q}_0$ from error subspaces). Depending on a  combination of the two errors, the logical qubit $|\alpha,\beta \rangle_{\rm L}$ may be affected by logical errors $X_{\rm L}$, $Y_{\rm L}$, or $Z_{\rm L}$. The solid-line arrows illustrate a scenario of errors $X_1$ and $X_3$, leading to logical error $X_{\rm L}$; the dashed-line arrows correspond to errors $Z_4$ and $Z_1$, leading to $Z_{\rm L}$.
}
\label{fig:continuous-scheme}
\end{figure}

A logical error may occur when the second single-qubit error moves the four-qubit state back to the subspace ${\cal Q}_0$ before the approximate cross-correlators $C_{12}$ and $C_{34}$ report an error. An example of evolution leading to a logical error is illustrated by solid-line arrows in Fig.\ \ref{fig:continuous-scheme}; another example is illustrated by dashed-line arrows. The thick-line circle illustrates evolution of the gauge qubit within the code space ${\cal Q}_0$ -- see Eq.\ (\ref{psi-a-b}). An error $X_i$ in $i$th qubit occurring at time $t_1$ instantaneously moves the state to the subspace ${\cal Q}_X$, and the state then continues to evolve due to measurement (errors $X_3$ or $X_4$ lead to evolution within a 2-dimensional subspace, which is different from the subspace after errors $X_1$ or $X_2$). Similarly, errors $Y_i$ would move the state from ${\cal Q}_0$ to ${\cal Q}_Y$, and $Z_i$ errors move from ${\cal Q}_0$ to ${\cal Q}_Z$, with generally different 2-dimensional subspaces for different $i$. (Figure \ref{fig:continuous-scheme} will be discussed in more detail later.) Even though the states in error subspaces ${\cal Q}_{X,Y,Z}$ are distinguishable from legitimate states in ${\cal Q}_0$ via cross-correlators, the error will be reported only after (on average) the response time $T_{\rm R}$. Therefore, if the second error occurs at time $t_1<t_2<t_1+T_{\rm R}$ and moves the state back to ${\cal Q}_0$, then both errors will (most likely) remain undetected. (Various combinations of the two errors at times $t_1$ and $t_2$ lead to various logical errors, which correspond to the classification in Sec.\ \ref{sec:2-qubit-errors}.)

As a result, the rate of logical errors is proportional to the response time $T_{\rm R}$, leading to formulas somewhat similar to Eqs.\ (\ref{proj-X-Mark})--(\ref{proj-Z-Mark}) for the projective case, with $\Delta t$ replaced by $T_{\rm R}$. We will see that $T_{\rm R}$ is comparable to the correlator integration time $T_{\rm c}$, which is typically chosen an order of magnitude larger than the collapse timescale $\tau_{\rm m}$. Thus, the operation characteristics of the Bacon-Shor code with continuous and projective measurements are generally similar to each other when $\tau_{\rm m}$ is comparable to $10^{-1} \Delta t$.

Encoding of the logical qubit can still be done using the gate operations $U_{\rm enc}$ in Eq.\ (\ref{U-enc}) and Fig.\ \ref{fig:circuit}, producing the state $|z+\rangle$. For the decoding, we can stop continuous measurement of $X_{12}$ and $X_{34}$, so that after several $\tau_{\rm m}$ we have essentially measured $Z_{13}$ and $Z_{24}$ projectively, and after that the decoding is the same as for the projective-measurement case (applying either $U_{\rm enc}^\dagger$ or $U_{\rm enc}^\dagger X_1X_2$).

In the following sections we present derivations and quantitative results for what was discussed in this overview.

\subsection{General evolution due to measurement}\label{sec:cont-gen-evolution}

It is easy to understand physics of evolution due to continuous measurement using the quantum Bayesian approach \cite{Korotkov-1999-2001,Korotkov-2003,Korotkov-2002ent}. For simplicity let us start with measuring only one gauge operator $G_k$ by an ideal (quantum-limited) detector and use the wavefunction language. An arbitrary four-qubit wavefunction at time $t$ can be represented as
    \be
    |\psi (t)\rangle =c_+  |\psi_+\rangle + c_-  |\psi_-\rangle,
    \label{psi-pm}\ee
where $|\psi_\pm\rangle$ are the normalized components belonging to the subspaces with eigenvalues $\pm 1$ (i.e., $G_k |\psi_\pm\rangle =\pm |\psi_\pm \rangle$), and $c_\pm$ are complex coefficients, $|c_+|^2+|c_-|^2=1$. The (inverse) measurement strength can be characterized by time $\tau_k$ needed to distinguish the states in the two subspaces with signal-to-noise ratio of 1 (the standard though misleading name for $\tau_k$ is ``measurement time''; we will also use notation $\tau_{\rm m}$ when all measurements have equal strength). The detector (one-channel) output $I_k(t)$ is assumed to contain white noise (Markovian case), and we also assume that the state evolves only due to (quantum non-demolition) measurement of $G_k$. Then the normalized output signal $\bar{I}_k$ averaged over time $\delta t$ between $t$ and $t+\delta t$ obviously has the probability distribution consisting of two Gaussians, \begin{eqnarray}
&&    P(\bar{I}_k)= |c_+|^2 P_+(\bar{I}_k) + |c_-|^2 P_- (\bar{I}_k),
    \label{P(I)}\\
&& P_\pm (\bar{I}_k) = \frac{1}{\sqrt{2\pi D}} \, \exp [-(\bar{I}_k \mp 1)^2/2D], \,\,\, D=\frac{\tau_k}{\delta t}, \qquad
    \label{P-pm(I)} \\
&& \bar{I}_k =\frac{1}{\delta t} \int_{t}^{t+\delta t} I_k(t')\, dt'.
    \end{eqnarray}
Note that the Gaussians are centered at $\pm 1$, which are the average signals (eigenvalues) for states $|\psi_\pm\rangle$, and the variance $D$ decreases with a longer averaging time $\delta t$.

The simplest model of the evolution due to measurement with a particular result $\bar{I}_k$ includes only the quantum Bayesian update (purely quantum or ``informational'' back-action \cite{Korotkov-2011}),
    \be
    |\psi (t+\delta t)\rangle = \frac{\sqrt{P_+(\bar{I}_k)} \, c_+  |\psi_+\rangle +  \sqrt{P_-(\bar{I}_k)} \, c_-  |\psi_-\rangle }{\rm Norm},
    \label{Bayesian-update}\ee
which is consistent with the classical Bayes rule for probabilities (see derivation in \cite{Korotkov-1999-2001}).

A somewhat extended model \cite{Korotkov-2003,Korotkov-2011} also includes ``classical'' phase back-action,
   \begin{eqnarray}
   && |\psi (t+\delta t)\rangle = \left[  \exp[-i({\cal K}_k \bar{I}_k +\varepsilon_k)\, \delta t]  \sqrt{P_-(\bar{I}_k)} \, c_-  |\psi_-\rangle \right.
    \nonumber \\
   && \hspace{2.2cm}    \left.  +  \sqrt{P_+(\bar{I}_k)} \, c_+ |\psi_+\rangle \right] /{\rm Norm}
     ,
    \label{phase-back-action}\end{eqnarray}
where coefficient ${\cal K}_k$ characterizes the phase back-action proportional to the output signal, and $\varepsilon_k$ is the effective energy shift between the subspaces due to measurement.
The phase back-action, for example, is important in phase-sensitive circuit QED measurement \cite{Gambetta-2008,Korotkov-2011} when a non-optimal quadrature is amplified.

While Eqs.\ (\ref{P(I)})--(\ref{phase-back-action}) describe the main physics of the continuous measurement of $G_k$ operators (in Markovian case), we often need to apply a few additional technical steps \cite{Korotkov-1999-2001,Korotkov-2003}. First, we can easily convert Eqs.\ (\ref{P(I)})--(\ref{phase-back-action}) into the language of density matrix, viewing it as a mixture of wavefunctions. Second, we generalize these equations to a non-ideal detector \cite{Korotkov-nonid} by adding a classical noise at the output and a noise causing decoherence between the subspaces (with a possible correlation between them). Averaging over these noises (which cannot be separately monitored by an observer) leads to decoherence. Third, we can convert the description (\ref{P(I)})--(\ref{phase-back-action}) with finite $\delta t$ into a differential form (infinitesimal $\delta t$). Note that for measurement of only one operator (with no other evolution), $\delta t$ can be arbitrarily long; however, when we simultaneously measure non-commuting observables, $\delta t$ should be short, so that the state change due to other evolution within $\delta t$ can be neglected.

For infinitesimal $\delta t$, Eqs.\ (\ref{P(I)}) and (\ref{P-pm(I)}) can be replaced with the single Gaussian with shifted center,
    \be
    P(\bar{I}_k) = (2\pi D)^{-1/2} \, \exp \{-[\bar{I}_k -{\rm Tr}(G_k\rho)]^2/2D\},
    \ee
and therefore the output signal $I_k(t)$ can be written as
    \be
    I_k(t) ={\rm Tr}[G_k \rho (t)] + \sqrt{\tau_k}\, \xi_k(t),
    \label{Il(t)}\ee
where $\rho(t)$ is the four-qubit density matrix and $\xi_k(t)$ is the white noise with correlator
    \be
\langle \xi_k(t)\,\xi_{k'}(t')\rangle =\delta_{kk'}\delta(t-t'),
    \label{xi-corr}\ee
i.e., integral of $\xi_k$ is the Wiener process, and there is no correlation between noises in different detectors. Note that Eq.\ (\ref{Il(t)}) remains valid for a non-ideal detector because $\tau_k$ is defined via the total noise (this is the distinguishability time for an observer).

When converting the evolution equations (\ref{Bayesian-update}) or (\ref{phase-back-action}) into the differential form, it is necessary to pay attention to the definition of the derivative \cite{Korotkov-2003}, since we are dealing with noise, and equations are nonlinear. The most widely used definitions are \cite{Oksendal-book} $\dot{f}(t)=\lim_{\Delta t\rightarrow 0} [f(t+\Delta t/2)-f(t-\Delta t/2)]/\Delta t$ (so-called Stratonovich form) and $\dot{f}(t)=\lim_{\Delta t\rightarrow 0} [f(t+\Delta t)-f(t)]/\Delta t$ (so-called It\^o form). The Stratonovich form is more physically intuitive since it preserves the usual calculus; the It\^o form modifies the usual calculus (requiring It\^o calculus), but makes averaging easy.

In this way from Eq.\ (\ref{phase-back-action}) we can derive the following evolution equation in the It\^o form:
\begin{eqnarray}
&& \dot{\rho} = \frac{i}{2}\left({{\cal K}_k}{\sqrt{\tau_k}}\, \xi_k+ \varepsilon_k \right) [ G_k,\rho] + \frac{\Gamma_k}{2} (G_k\rho G_k-\rho)
 \nonumber \\
&& \hspace{1.2cm}  + \frac{\xi_k}{2\sqrt{\tau_k}}\left(G_k\rho + \rho G_k - 2\rho\,{\rm Tr}[G_k\rho ]\right),
\label{evolution-rho-1}\end{eqnarray}
where the noise $\xi_k(t)$ is the same as in Eq.\ (\ref{Il(t)}), the measured observable is Hermitian, $G_k^\dagger =G_k$, with $G_k^\dagger G_k=\openone$ (if $G_k^\dagger G_k\neq \openone$, then the last term on the first line should be replaced with the Lindblad form), and
the effective ensemble dephasing $\Gamma_k$ satisfies inequality \cite{Korotkov-2003} $\Gamma_k\geq 1/2\tau_k+ {\cal K}_k^2\tau_k/2$ (the notation $\Gamma_k$ should not be confused with the previous notation $\Gamma_i^{(X,Y,Z)}$ for the error rates). Note that Eq.\ (\ref{evolution-rho-1}) can also be derived using the theory of quantum trajectories \cite{Wiseman-Milburn-1993,Carmichael-1993,Wiseman-Milburn-book}.

When several (non-commuting) gauge operators $G_k$ are continuously measured at the same time, the density matrix evolution (\ref{evolution-rho-1}) due to each measurement should be simply added up \cite{Ruskov-2010} (this relates to the fact that infinitesimal evolutions essentially commute with each other). Also adding the Lindblad evolution (\ref{Lindblad-1}) described by error operators $E_i$, we obtain the overall evolution
\begin{eqnarray}
&& \hspace{-0.2cm} \dot{\rho} = \sum\nolimits_k\left[ \frac{i}{2}\left({{\cal K}_k}{\sqrt{\tau_k}}\, \xi_k+ \varepsilon_k \right) [ G_k,\rho] + \frac{\Gamma_k}{2} (G_k\rho G_k-\rho) \right.
 \nonumber \\
&& \hspace{1.2cm} \left.
+ \frac{\xi_k}{2\sqrt{\tau_k}}\left(G_k\rho + \rho G_k - 2\rho\,{\rm Tr}[G_k\rho ]\right)\right]
 \nonumber \\
 &&\hspace{0.4cm}
+ \sum\nolimits_{i,E}\Gamma_i^{(E)}{\cal L}[E_i]\rho .
\label{evolution-rho}\end{eqnarray}

The quantum efficiency of each detector can be defined in two ways \cite{Korotkov-2003,Korotkov-nonid}:
    \be
    \eta_k = \frac{1}{2\Gamma_k \tau_k}, \,\,\,  \tilde{\eta}_k = \frac{1+{\cal K}_k^2\tau_k^2}{2\Gamma_k \tau_k}, \,\,\, \eta_k\leq \tilde{\eta_k}\leq 1.
    \ee
The first definition relates ensemble decoherence with the rate of distinguishing the subspaces, while the second definition compares ensemble decoherence with its information-related part, including the phase back-action.

If all detectors are ideal in the sense $\tilde\eta_k=1$, then the evolution (\ref{evolution-rho}) can also be described with a wavefunction (if initial state is pure and decoherence ${\cal L}[E_i]$ is unraveled in the ``jump/no-jump'' way), i.e., the measurement evolution description (\ref{phase-back-action}) with small $\delta t$ is fully sufficient. When detectors are non-ideal, sometimes it is also possible to work with wavefunctions (that greatly simplifies analysis) in the following way. A non-ideal detector can be thought of as an ideal detector with an uncorrelated extra noise at the output \cite{Korotkov-nonid}, so that Eq.\ (\ref{Il(t)}) contains an extra (classical) noise term, while Eq.\ (\ref{evolution-rho}) is governed only by the quantum part of the noise, with $\tau_k$ corresponding to the ideal part of the detector. Separation of the output noise into the quantum and classical part is not possible for an observer, but we may pretend that it is possible for a ``Supreme Being'', who, therefore, can monitor the wavefunction evolution. Thus, we can use predictions for ideal detectors, while remembering about extra noise at the output. This logical trick will be quite useful in our analysis.

For numerical simulations in full 16-dimensional Hilbert space, it is sufficient to use evolution equation (\ref{evolution-rho}) and Eq.\ (\ref{Il(t)}) for the output signal (in practice, instead of working with Wiener processes, it is usually better to use explicit quantum Bayesian procedure \cite{Korotkov-1999-2001}). However, these Monte Carlo simulations are numerically expensive and also not quite suitable to obtain analytical results.
For analytics it is easier to discuss separate evolutions in the four subspaces ${\cal Q}_{0,X,Y,Z}$, with jumps between them caused by single-qubit errors. This is what we will do next.

\subsection{Evolution without errors}\label{sec:cont-no-error-evol}

Let us prepare initial encoded state $|z+\rangle$ [Eq.\ (\ref{z+})] and start continuous measurement of four gauge operators $G_k$, assuming no evolution due to environment (only due to measurement). First, let us show that the four-qubit state remains within the subspace spanned by $|z+\rangle$ and $|z-\rangle$. This can be shown using either Eq.\ (\ref{phase-back-action}) or Eq.\ (\ref{evolution-rho}). For the proof using Eq.\ (\ref{evolution-rho}), note that
all operators $G_k$ applied to $|z+\rangle$ or $|z-\rangle$, produce states within the subspace spanned by $|z\pm\rangle$. Therefore, if the four-qubit state $\rho$ is within the subspace generated by $|z\pm\rangle$ (i.e.\, spanned by $|z+\rangle \langle z+|$, $|z-\rangle \langle z-|$, $|z+\rangle \langle z-|$, and $|z-\rangle \langle z+|$), then the right-hand side of Eq.\ (\ref{evolution-rho}) is also within this subspace, so that the state remains in this subspace during the evolution due to measurement.

It is also instructive to use Eq.\ (\ref{phase-back-action}) assuming ideal detectors and show explicitly that the evolving state is described by the wavefunction (\ref{psi-a-b}) (additional output noise of non-ideal detectors can be added later). For measurement of operator $G_3=Z_{13}$ or $G_4=Z_{24}$ and quantum state $|\psi\rangle =a \, |z+\rangle +b \, |z-\rangle$ [Eq. (\ref{psi-a-b})], the eigenvectors $|\psi_{\pm}\rangle$ in Eq.\ (\ref{psi-pm}) are simply $|z\pm\rangle$. Then the evolution (\ref{phase-back-action}) changes coefficients $a$ and $b$, still preserving the form (\ref{psi-a-b}). For measurement of operator $G_1=X_{12}$ or $G_2=X_{34}$, the eigenvectors $|\psi_{\pm}\rangle$ in Eq.\ (\ref{psi-pm}) are $|x\pm\rangle$, which are linear combinations of $|z\pm\rangle$ [Eq.\ (\ref{x-pm-z-pm})]. Therefore, the evolution (\ref{phase-back-action}) still keeps the state within the 2-dimensional subspace (\ref{psi-a-b}). So, measurement of all gauge operators only changes the gauge qubit $|a,b\rangle_{\it g}$ in Eq.\ (\ref{psi-a-b}), while not affecting the logical qubit $|\alpha ,\beta\rangle_{\rm L}$, which defines the basis $|z\pm\rangle$. Detector non-ideality leads to an imperfect knowledge of $a$ and $b$ for an observer (while they are perfectly known to the ``Supreme Being''), therefore, for an observer the gauge qubit states (\ref{psi-a-b}) are mixed, producing density matrix $\rho_{\it g}$ for the gauge qubit, while the logical qubit $|\alpha ,\beta\rangle_{\rm L}$ is not disturbed.

It is easy to see that within the gauge qubit subspace spanned by $|z+\rangle$ and $|z-\rangle$, continuous measurement of $G_3$ and $G_4$ is the usual $Z$-measurement of the gauge qubit, while $G_1$ and $G_2$ correspond to continuous $X$-measurement of the gauge qubit. Therefore, we have simultaneous $X$ and $Z$ measurement of a qubit, which was described theoretically in Ref.\ \cite{Ruskov-2010} and realized experimentally in Ref.\ \cite{Hacohen-Courgy-2016}. Using results of \cite{Ruskov-2010} and adding the phase back-action, from Eq.\ (\ref{evolution-rho}) we obtain the following explicit equations for the evolution of the Bloch-sphere components of the gauge qubit density matrix $\rho_{\it g}=(\openone +x_g \sigma_x+y_g\sigma_y +z_g \sigma_z)/2$ (in It\^o form):
\begin{eqnarray}
&& \hspace{-0.3cm} \dot x_g = (1 - x_g^2)\left(\frac{\xi_1}{\sqrt{\tau_1}} + \frac{\xi_2}{\sqrt{\tau_2}}\right) + \left({\cal K}_3\tau_3y_g - x_gz_g \right)\frac{\xi_3}{\sqrt{\tau_3}}
          \nonumber \\
&& + \left({\cal K}_4\tau_4 y_g - x_gz_g \right)\frac{\xi_4}{\sqrt{\tau_4}}
 -(\Gamma_3 + \Gamma_4)x_g + (\varepsilon_3+\varepsilon_4)y_g ,
    \nonumber \\
     \label{xg-1}\\
&& \hspace{-0.3cm} \dot y_g =  \left({\cal K}_1\tau_1z_g - x_gy_g \right)\frac{\xi_1}{\sqrt{\tau_1}} +
\left({\cal K}_2\tau_2z_g - x_gy_g \right)\frac{\xi_2}{\sqrt{\tau_2}}
    \nonumber \\
&& - \left({\cal K}_3\tau_3x_g + y_gz_g \right)\frac{\xi_3}{\sqrt{\tau_3}}  - \left({\cal K}_4\tau_4x_g + y_gz_g \right)\frac{\xi_4}{\sqrt{\tau_4}}
    \nonumber \\
&& -(\Gamma_1 + \Gamma_2 + \Gamma_3 + \Gamma_4)y_g  - (\varepsilon_3 + \varepsilon_4)x_g + (\varepsilon_1 + \varepsilon_2)z_g ,
    \nonumber \\
    \label{yg-1}\\
&& \hspace{-0.3cm} \dot z_g =  ( 1 - z_g^2) \left (\frac{\xi_3}{\sqrt{\tau_{3}}} + \frac{\xi_4}{\sqrt{\tau_{4}}}\right)  - \left({\cal K}_1\tau_1 y_g + x_gz_g \right)\frac{\xi_1}{\sqrt{\tau_{1}}}
    \nonumber\\
&& - \left({\cal K}_2\tau_2 y_g + x_gz_g \right) \frac{\xi_2}{\sqrt{\tau_{2}}} -(\Gamma_1 + \Gamma_2)z_g   - (\varepsilon_1+\varepsilon_2)y_g  ,
    \nonumber\\ \label{zg-1}
\end{eqnarray}
while the measurement output signals are
    \begin{eqnarray}
    && \hspace{-1.0cm} I_1=I_{X12}= x_g + \sqrt{\tau_1}\,\xi_1, \,\,\,  I_2=I_{X34}= x_g + \sqrt{\tau_2}\,\xi_2,
    \label{I1,2}\\
    && \hspace{-1.0cm} I_3=I_{Z13}= z_g + \sqrt{\tau_3}\,\xi_3, \,\,\,  I_4=I_{Z24}= z_g + \sqrt{\tau_4}\,\xi_4.
    \label{I3,4}\end{eqnarray}

Note that in deriving Eqs.\ (\ref{xg-1})--(\ref{I3,4}) we assumed that the four-qubit state is fully in the subspace ${\cal Q}_0$, and in the basis of four vectors $|\phi_j\rangle$ [Eqs. (\ref{phi-1})--(\ref{phi-4})] the density matrix is a direct product of the logical and gauge qubit states, i.e.,
    \be
    \rho_{{\cal Q}} = \rho_{\rm L} \otimes \rho_{\it g} = \left(\begin{array}{cc} \frac{1+z_{\rm L}}{2}\times \rho_{\it g}  & \frac{x_{\rm L}-iy_{\rm L}}{2}\times \rho_{\it g}  \\  \frac{x_{\rm L}+iy_{\rm L}}{2}\times \rho_{\it g}  & \frac{1-z_{\rm L}}{2}\times \rho_{\it g}  \end{array} \right)_{4\times 4} ,
    \label{rho-4x4}\ee
where $x_{\rm L}, y_{\rm L}, z_{\rm L}$ are components of the logical qubit state $\rho_{\rm L}$. The measurement does not affect the logical qubit state, $\dot{x}_{\rm L}=\dot{y}_{\rm L}=\dot{z}_{\rm L}=0$.

We will mostly consider the special case when there is no phase back-action and all four measurements have equal measurement strength and corresponding ensemble dephasing,
    \be
    \tau_k=\tau_{\rm m}, \,\,\, \Gamma_k=\Gamma_{\rm m}, \,\,\,  {\cal K}_k=0, \,\,\, \varepsilon_k=0,\,\,\, k=1,\dots 4 .
    \label{uniform-4}\ee
In this case the evolution equations (\ref{xg-1})--(\ref{zg-1}) simplify,
\begin{eqnarray}
&& \hspace{-0.3cm} \dot x_g = (1 - x_g^2)\,\frac{\xi_1+\xi_2}{\sqrt{\tau_{\rm m}}}  - x_gz_g \frac{\xi_3+\xi_4}{\sqrt{\tau_{\rm m}}}
 -2\Gamma_{\rm m} x_g ,
        \label{xg-2}\\
&& \hspace{-0.3cm} \dot y_g =  - x_gy_g \frac{\xi_1+\xi_2}{\sqrt{\tau_{\rm m}}}
 - y_gz_g \frac{\xi_3+\xi_4}{\sqrt{\tau_{\rm m}}}
 -4\Gamma_{\rm m} y_g ,
    \label{yg-2}\\
&& \hspace{-0.3cm} \dot z_g =  ( 1 - z_g^2) \, \frac{\xi_3+\xi_4}{\sqrt{\tau_{\rm m}}}  - x_gz_g \frac{\xi_1+\xi_2}{\sqrt{\tau_{\rm m}}} -2\Gamma_{\rm m} z_g .
    \label{zg-2}
\end{eqnarray}
We see that the component $y_g$ exponentially decreases towards zero on the timescale of $(4\Gamma_{\rm m})^{-1}$, while in the $xz$-plane the evolution is isotropic (this can be seen by considering linear combinations of $x_g$ and $z_g$). In particular, in the ideal case when $\Gamma_{\rm m}=1/2\tau_{\rm m}$ and the initial state is $|z+\rangle$, the evolution can be described by a simple uniform diffusion of the wavefunction (\ref{psi-a-b}) along the great circle of the Bloch sphere \cite{Ruskov-2010,Hacohen-Courgy-2016}, so that coefficients $a(t)$ and $b(t)$ in Eq.\ (\ref{psi-a-b}) are real. In a non-ideal case, $\Gamma_{\rm m} > 1/2\tau_{\rm m}$, the evolution can still be viewed in this way for the ``Supreme Being'', as discussed above.

For non-equal strength of four measurements, evolution of the gauge qubit state is the diffusion with the state-dependent diffusion coefficient and also the drift along the Bloch sphere. In the presence of phase back-action, the coefficients $a$ and $b$ in Eq.\ (\ref{psi-a-b}) are necessarily complex, so the whole Bloch sphere is involved in evolution. This complicates the analysis, but general picture remains the same: measurement causes continuous evolution of the gauge qubit, without disturbing the logical qubit.

\subsection{Measurement evolution within error subspaces}\label{sec:cont-error-evol}

Suppose the error $X_1$ has occurred. Immediately after this error, the four-qubit state still has the form (\ref{rho-4x4}) with the same gauge and logical qubit states, but with the basis vectors $|\phi_j\rangle$ [Eqs.\ (\ref{phi-1})--(\ref{phi-4})] replaced with $X_1|\phi_j\rangle$. In other words, the $4\times 4$ matrix (\ref{rho-4x4}) moves to the different block of the full $16\times 16$ matrix. After that, the continuous measurement of gauge operators $G_k$ again leads to an evolution of the gauge qubit state without affecting the logical qubit. The only difference compared with the previous section is that in the basis of vectors $X_1|\phi_j\rangle$, the operator $G_3=Z_{13}$ has the opposite eigenvalue, $Z_{13} (X_1|\phi_j\rangle)=-(X_1|\phi_j\rangle)$, while eigenvalues for other three gauge operators are still $+1$. This means that $G_3$ now measures the gauge qubit along $-Z$ axis instead of $Z$ axis.

Therefore, evolution equations (\ref{xg-1})--(\ref{I3,4}) should be changed within the error subspace ${\cal Q}_X$, in particular, now ${\rm Tr} [G_3\rho] = -z_g$. If we write the output signal $I_3(t)$ as $I_3=-z_g +\sqrt{\tau_3}\, \xi_3$ instead of Eq.\ (\ref{I3,4}), then we also need to replace $\xi_3$ with $-\xi_3$ in Eqs.\ (\ref{xg-1})--(\ref{zg-1}). However, it is easier to flip the sign in the definition of $\xi_3$, so that
    \be
    I_3 =I_{Z13} = -(z_g +\sqrt{\tau_3}\, \xi_3),
    \label{I-3-X}\ee
then the evolution equations (\ref{xg-1})--(\ref{zg-1}) do not change. This mapping can be interpreted as being due to the transformation $X_1 G_3 X_1 =-G_3$ (somewhat similar to the Heisenberg picture, in which the error-mapping $\rho \to E_i\rho E_i$ is instead applied to the measured operators).

Thus, for a four-qubit state in the error subspace ${\cal Q}_X$ (still assuming a direct product of the gauge and logical qubit states), the dynamics due to continuous measurement is the same as in the subspace ${\cal Q}_0$, except Eq.\ (\ref{I-3-X}) for the signal $I_3(t)$ replaces Eq.\ (\ref{I3,4}).

A similar reasoning shows that after an error $Z_1$, the dynamics due to measurement in the subspace ${\cal Q}_Z$ in the basis $Z_1|\phi_j\rangle$ is still described by Eqs.\  (\ref{xg-1})--(\ref{I3,4}), except now
    \be
    I_1 = I_{X12} = -(x_g+\sqrt{\tau_1}\, \xi_1).
    \label{I-1-Z}\ee
Finally, after an error $Y_1$, the dynamics in the subspace ${\cal Q}_Y$ in the basis $Y_1|\phi_j\rangle$ is described by Eqs.\  (\ref{xg-1})--(\ref{I3,4}) with the change for both $I_1$ and $I_3$,
    \be
    I_1 = -(x_g+ \sqrt{\tau_1}\, \xi_1), \,\,\,
     I_3 = -(z_g +\sqrt{\tau_3}\, \xi_3)  .
    \label{I-13-Y}\ee

Note that we intentionally considered only errors in the first qubit ($X_1$, $Y_1$, $Z_1$) because we use the 16-dimensional Hilbert space basis consisting of $|\phi_j\rangle$, $X_1|\phi_j\rangle$,  $Y_1|\phi_j\rangle$, and  $Z_1|\phi_j\rangle$. The mapping between the states due to errors in other qubits will be considered next, while the evolution in the error subspaces due to measurement after these errors is the same as already discussed.

\subsection{Mapping between subspaces due to single-qubit errors}\label{sec:cont-error-mapping}

As discussed above, an error $X_1$ occurring at time $t$, by definition does not change the logical and gauge qubit states, $\rho_{\it g}(t+0)=\rho_{\it g}(t-0)$, $\rho_{\rm L}(t+0)=\rho_{\rm L}(t-0)$, and only moves the four-qubit state from ${\cal Q}_0$ to ${\cal Q}_X$. To analyze the effect of the error $X_2$ (instead of $X_1$), we compare it with the effect of $X_1$. It is easy to see that $X_2$ acting on the basis $|\phi_j\rangle$ produces the same states as $X_1$ with additional exchange: $|\phi_1\rangle\leftrightarrow |\phi_2\rangle$ and $|\phi_3\rangle\leftrightarrow |\phi_4\rangle$. Therefore, $X_2$ acting on a state (\ref{psi-a-b}) produces the same state as $X_1$, but with exchanged gauge qubit coefficients, $a\leftrightarrow b$. Consequently, for a more general initial state (\ref{rho-4x4}), the application of $X_2$ produces the same state as application of $X_1$ and additional $X$-operator for the gauge qubit state, which we denote as $X_{\rm G}$.

Thus, we associate effect of error $X_2$ (acting on a state within ${\cal Q}_0$) with the error $X_1$ and gauge-qubit operation $X_{\rm G}$. In a similar way we find that $X_3$ acts on the basis $|\phi_j\rangle$ as $X_1$ with additional exchange $|\phi_1\rangle\leftrightarrow |\phi_3\rangle$ and $|\phi_2\rangle\leftrightarrow |\phi_4\rangle$, therefore acting on the state (\ref{psi-a-b}) as $X_1$ with exchange $\alpha \leftrightarrow \beta$. Thus, effect of error $X_3$ is the same as for the error $X_1$ and logical-qubit $X$-operation, which we denote as $X_{\rm L}$. Similarly, we find that $X_4$ is equivalent to $X_1$ with additional operations $X_{\rm G}X_{\rm L}$ on both gauge and logical qubits.

Similarly, we can find the effect of the errors $Z_i$ comparing them with $Z_1$, and effect of the errors $Y_i$ in comparison with $Y_1$. The result is the following correspondence:
    \begin{eqnarray}
    && \hspace{-0.3cm} X_2\leftrightarrow X_1\times X_{\rm G}, \,\,\,   X_3\leftrightarrow X_1\times X_{\rm L}, \,\,\,
       X_4\leftrightarrow X_1\times X_{\rm G}X_{\rm L},
\nonumber \\
       && \hspace{-0.3cm} Z_2\leftrightarrow Z_1\times Z_{\rm L}, \,\,\,   Z_3\leftrightarrow Z_1\times Z_{\rm G}, \,\,\,
       Z_4\leftrightarrow Z_1\times Z_{\rm G}Z_{\rm L},
       \nonumber \\
 && \hspace{-0.3cm} Y_2\leftrightarrow Y_1\times (-X_{\rm G} Z_{\rm L}), \,\,\,   Y_3\leftrightarrow Y_1\times (-Z_{\rm G}X_{\rm L}),
    \nonumber \\
 && \hspace{2.7cm} Y_4\leftrightarrow Y_1\times (-X_{\rm G}Z_{\rm G} X_{\rm L} Z_{\rm L}) .
    \label{error-correspondence}\end{eqnarray}
This mapping is illustrated in Fig.\ \ref{fig:continuous-scheme}.
Note that if a state of the direct-product form (\ref{rho-4x4}) is returned from an error subspace to ${\cal Q}_0$ by another single-qubit error, then the same correspondence applies, as can be easily shown using relations $X_i=X_1X_iX_1$, $Y_i=Y_1Y_iY_1$, and $Z_i=Z_1Z_iZ_1$.

\subsection{Logical two-qubit errors}\label{sec:cont-logical-errors}

\subsubsection{Uncorrelated Markovian errors}\label{sec:Markovian-cont}

We can now discuss the mechanism of logical errors, using the model of uncorrelated Markovian single-qubit errors introduced in Sec.\ \ref{sec:Markovian-proj}. As an example, let us assume that the first single-qubit error in the procedure is $X_1$ and it occurs at time $t_1$ (this error is indicated by the upper solid-line arrow in Fig.\ \ref{fig:continuous-scheme}). The state evolution between $t=0$ (preparation of the state $|z+\rangle$) and $t_1$ is evolution of the gauge qubit (illustrated by the thick-line circle in Fig.\ \ref{fig:continuous-scheme}), without change of the logical qubit state. The error $X_1$ moves the state from ${\cal Q}_0$ to ${\cal Q}_X$ without change of the logical and gauge qubit states, and after that the gauge qubit continues to evolve (not affecting the logical qubit). Monitoring of time-integrated correlators constructed from the output signals $I_k (t)$ (discussed later) is supposed to report that the error has occurred; however, it takes some time to find this out, so that the error is reported (on average) at time $t=t_1+T_{\rm R}$, where the {\it average response time} $T_{\rm R}$ will be calculated later. If another single-qubit error, for example $X_3$, occurs at time $t_2$ within the interval $[t_1, t_1+T_{\rm R}]$ and moves the state back to ${\cal Q}_0$, then the error will (most likely) not be reported, since after $t_2$ the correlators are normal again. The state is returned to ${\cal Q}_0$, but it is not returned to the proper 2-dimensional subspace (see Fig.\ \ref{fig:continuous-scheme}) because $X_3$ error applied $X_{\rm L}$ operation to the logical qubit, as follows from Eq.\ (\ref{error-correspondence}) (two $X_1$ errors cancel out each other, and there is also an unimportant gauge qubit evolution between $t_1$ and $t_2$). Thus a logical $X$-error is produced [note the $X_1X_3$ combination in Eq.\ (\ref{log-X})].

To analyze the effect of different combinations of single-qubit errors, we can use the correspondence relations (\ref{error-correspondence}). The $X_i$ error moves the legitimate state to one of two subspaces in ${\cal Q}_X$, either with or without $X_{\rm L}$ operation (see two circles in the second column in Fig.\ \ref{fig:continuous-scheme}), while possible application of $X_{\rm G}$ is not important. Similarly, $Z_i$ error moves the legitimate state to ${\cal Q}_Z$ either with or without $Z_{\rm L}$ (two circles in Fig.\ \ref{fig:continuous-scheme}) while possible $X_{\rm G}$ operation is not important. The errors $Y_i$ move the state from ${\cal Q}_0$ to four different subspaces [see Eq.\ (\ref{error-correspondence}) and Fig.\ \ref{fig:continuous-scheme}]. If the second error (occurring at $t_2$) does not bring the state to ${\cal Q}_0$, then an error will be detected; therefore, we are interested only in error combinations returning the state back to ${\cal Q}_0$. There are harmless combinations (i.e., $X_1X_1$, $X_1X_2$, etc.), which do not produce logical errors, and there are combinations producing three types of logical errors (see four circles in the left column in Fig.\ \ref{fig:continuous-scheme}). The logical errors illustrated in Fig.\ \ref{fig:continuous-scheme} are $X$-error due to $X_1$ and $X_3$ (solid-line arrows) and also $Z$-error due to $Z_4$ and $Z_1$ (dashed-line arrows).
Since the evolution of the gauge qubit due to single-qubit errors and due to measurement between $t_1$ and $t_2$ is not important for us, the logical error combinations obtained from Eq.\ (\ref{error-correspondence}) are the same as those discussed in Sec.\ \ref{sec:2-qubit-errors}. Note that the combinations of two errors occurring in the same qubit are either harmless or detectable.

The rates of logical $X$, $Y$, and $Z$ errors can be obtained by calculating the probability of the second error occurring within the response time $T_{\rm R}$ after the first error, and summing over the error combinations. It is important that in our discussed later  construction of averaged correlators, the response time $T_{\rm R}$ is the same for detecting states in all error subspaces (${\cal Q}_X$, ${\cal Q}_Y$, ${\cal Q}_Z$). In this case the calculation of logical error rates (assuming no detected errors) gives
    \begin{eqnarray}
    && \gamma_X = 2T_{\rm R} \left[(\Gamma_1^{(X)}+\Gamma_2^{(X)})(\Gamma_3^{(X)}+\Gamma_4^{(X)}) \right.
        \nonumber \\
    &&\hspace{1.8cm} \left. +\Gamma_{1}^{(Y)}\Gamma_{3}^{(Y)}+\Gamma_{2}^{(Y)}\Gamma_{4}^{(Y)} \right] .
    \label{cont-X-Mark}\\
 &&   \gamma_Y= 2T_{\rm R} \left[ \Gamma_{1}^{(Y)}\Gamma_{4}^{(Y)}+\Gamma_{2}^{(Y)}\Gamma_{3}^{(Y)} \right] ,
    \label{cont-Y-Mark} \\
    && \gamma_Z= 2T_{\rm R} \left[(\Gamma_1^{(Z)}+\Gamma_3^{(Z)})(\Gamma_2^{(Z)}+\Gamma_4^{(Z)}) \right.
    \nonumber \\
    &&\hspace{1.8cm} \left. +\Gamma_{1}^{(Y)}\Gamma_{2}^{(Y)}+\Gamma_{3}^{(Y)}\Gamma_{4}^{(Y)} \right] ,
    \label{cont-Z-Mark}\end{eqnarray}
where $\Gamma_i^{(X)}$, $\Gamma_i^{(Y)}$,  $\Gamma_i^{(Z)}$ are the rates of single-qubit errors (see Sec.\ \ref{sec:Markovian-proj}) and the factor of 2 is due to different sequences of the two errors. In general, the response times may be different for different error subspaces ($T_{R,X}$, $T_{R,Y}$, $T_{R,Z}$); in this case each product of error rates in Eqs.\ (\ref{cont-X-Mark})--(\ref{cont-Z-Mark}) should be multiplied by the corresponding response time.

In particular, for the depolarizing channel with $\Gamma_i^{(X)}=\Gamma_i^{(Y)}=\Gamma_i^{(Z)}=\Gamma_{\rm d}/3$ we have
    \be
     \gamma_X = \gamma_Z = \frac{4}{3}\,\Gamma_{\rm d}^2T_{\rm R}, \,\,\, \gamma_Y=\frac{4}{9}\,\Gamma_{\rm d}^2T_{\rm R},
    \ee
so that the total logical error rate (with no detected errors)
is
    \be
\gamma_{\rm L}=\frac{28}{9}\,\Gamma_{\rm d}^2T_{\rm R}.
    \label{gamma-L-depol-cont}\ee

Note that Eqs.\ (\ref{cont-X-Mark})--(\ref{gamma-L-depol-cont}) are similar to the results (\ref{proj-X-Mark})--(\ref{gamma-L-depol-proj}) for projective measurements if the half-cycle time $\Delta t$ is replaced with the response time $T_{\rm R}$. The similarity is not exact because of different ``effective response times'' for $Y_i$ errors compared with  $X_i$ and $Z_i$ errors in the projective case, while in the continuous case all response times are the same.

\subsubsection{Pure dephasing}\label{sec:dephasing-cont}

Following the logic used in Sec.\ \ref{sec:dephasing-proj}, we can apply Eqs.\ (\ref{cont-X-Mark})--(\ref{cont-Z-Mark}) to the case of pure dephasing of physical qubits with rates $\Gamma_{\varphi,i}$ by using the correspondence $\Gamma_i^{(Z)}=\Gamma_{\varphi,i}/2$. This gives the logical error rates
    \begin{eqnarray}
    && \gamma_X=0, \,\,\, \gamma_{Y}=0, \,\,\,
     \label{gamma-XY-dephasing}\\
     && \gamma_{Z}= T_{\rm R} \, (\Gamma_{\varphi,1}+\Gamma_{\varphi,3})(\Gamma_{\varphi,2}+\Gamma_{\varphi,4})/2,
    \label{gamma-Z-dephasing}\end{eqnarray}
in the case when no errors are detected by the procedure. For equal dephasing in all qubits, $\Gamma_{\varphi,i}=\Gamma_\varphi$, we have \be
    \gamma_{\rm L}=\gamma_Z= 2\Gamma_{\varphi}^2 T_{\rm R},
    \label{gamma-L-cont-deph}\ee
which corresponds to Eq.\ (\ref{gamma-L-proj-deph}) with $\Delta t$ replaced with $T_{\rm R}$.

\subsubsection{Energy relaxation} \label{sec:relax-cont}

Following the logic of Sec.\ \ref{sec:relax-proj}, let us analyze the effect of energy relaxation in  the physical qubits at zero temperature (amplitude damping) with rates $\mu_{i}\equiv 1/T_{1,i}$. The ``no-jump'' evolution with the Kraus operator $\openone-\sum_i (\mu_{i}\,\delta t/2)\, \sigma_{+,i}\sigma_{-,i}=\openone-\delta t\sum_i \mu_i (\openone -Z_i)/4$  for a short duration $\delta t$ produces detectable $Z_i$ errors with the rate on the order of $\mu_{i}^2/\Gamma_{\rm m}$, where $\Gamma_{\rm m}$ is the dephasing due to measurement. These errors can be neglected in comparison with ``jump'' errors, since we assume $\mu_{i}\ll \Gamma_{\rm m}$. With no detected $Z_i$-errors, measurement process self-corrects the state disturbed by the ``no-jump'' evolution.  Therefore, in the leading order we can completely neglect the ``no-jump'' evolution (see discussion in Sec.\ \ref{sec:relax-proj}).

The ``jumps'' (energy relaxation events) due to operators $\sigma_{-,j}=(X_i+\imath Y_i)/2$ with rates $\mu_{i}{\rm Tr}(\sigma_{+,i}\sigma_{-,i}\rho)$ lead to detectable errors, unless the second energy relaxation event occurs within the response time $T_{\rm R}$, leading to a logical error. The state evolution between the jumps can be described by general equation (\ref{evolution-rho}), but it cannot be easily described by Eqs.\ (\ref{xg-1})--(\ref{zg-1}) because these equations assume a direct product of gauge and logic qubits within only one error subspace, while the operator $(X_i+\imath Y_i)/2$ produces a superposition between subspaces ${\cal Q}_X$ and ${\cal Q}_Y$.
Most importantly, since measurement distinguishes between these subspaces, the state will be gradually collapsed into one of them within the timescale comparable to $\tau_{\rm m}$. Therefore, coherence between the two subspaces necessarily decays with a time constant comparable to $\tau_{\rm m}$ (one more reason for the decay of ensemble-averaged coherence is the difference between the gauge qubit evolutions within the two subspaces for the same output signal).
Since the time difference between the two relaxation events is typically comparable to $T_{\rm R}$ and since in our case (as will be discussed later) $T_{\rm R}$ is an order of magnitude larger than $\tau_{\rm m}$, we can neglect coherence between the subspaces.

After neglecting coherence between the subspaces, the calculation of the logical error rates is simple: the logical errors are due to independent two-qubit errors $X_jX_i$ and $Y_jY_i$, occurring within $T_{\rm R}$. Using the probability rate $\mu_i/2$ of the first jump in $i$th qubit, probability $\mu_j T_{\rm R}/2$ of the second jump in $j$th qubit within time $T_{\rm R}$, and probability $1/4$ each for the combinations $X_j X_i$ and $Y_jY_i$, we obtain the logical error rates
    \begin{eqnarray}
    && \gamma_X = \frac{T_{\rm R}}{8} \left[ (\mu_{1}+\mu _{2})(\mu_{3}+\mu_{4})
      + \mu_{1}\mu_{3} + \mu_{2}\mu_{4}\right] , \quad
        \label{relax-X-cont}\\
    && \gamma_Y = \frac{T_{\rm R}}{8}\, ( \mu_{1}\mu_{4} + \mu_{2}\mu_{3} ) ,
         \label{relax-Y-cont}\\
    && \gamma_Z = \frac{T_{\rm R}}{8} \left[\mu_{1}\mu_{2} + \mu_{3}\mu_{4}\right] ,
     \label{relax-Z-cont}\end{eqnarray}
where the error combinations come from Eqs.\ (\ref{log-X})--(\ref{log-Z}) or from Eq.\ (\ref{error-correspondence}). Note that in Eq.\ (\ref{relax-X-cont}) we show the products $\mu_{1}\mu_{3}$ and $\mu_{2}\mu_{4}$ twice to emphasize similarity with Eq.\ (\ref{cont-X-Mark}).

The corresponding total logical error rate (with no detected errors) is
    \be
     \gamma_{\rm L}= \frac{T_R}{8}\left [ 2(\mu_{1}+\mu_{2})(\mu_{3} +\mu_{4}) + \mu_{1}\mu_{2} + \mu_{3}\mu_{4}   \right] .
    \ee

\subsection{Cross-correlators}\label{sec:cont-correlators}

We have found the rate of logical errors for a given response time $T_{\rm R}$, but we have  not calculated $T_{\rm R}$ yet. We have also not calculated the termination rate. Moreover, we have not yet discussed quantitatively how we can monitor the error syndrome.

As discussed in Sec.\ \ref{sec:overview}, the general idea is that in the subspace ${\cal Q}_0$ and for the direct-product state (\ref{rho-4x4}), the operators $G_1=X_{12}$ and $G_2=X_{34}$ both measure $X$-component of the gauge qubit; therefore the corresponding noisy outputs $I_{1}(t)$ and $I_{2}(t)$ should be positively correlated. Similarly, for a state within ${\cal Q}_0$ the outputs $I_3(t)=I_{Z13}(t)$ and $I_4(t)=I_{Z24}(t)$ are also positively correlated. However, for a direct-product state within subspace ${\cal Q}_X$, the operator $G_3=Z_{13}$ measures the gauge qubit along axis $-Z$, while $G_4=Z_{24}$ measures it along $Z$-axis; therefore, the noisy outputs $I_{3}(t)$ and $I_{4}(t)$ should be negatively correlated. Similarly, within ${\cal Q}_Z$ the outputs $I_{1}(t)$ and $I_{2}(t)$ should be negatively correlated, while within ${\cal Q}_Y$ both pairs of the output signals should be negatively correlated. By monitoring the cross-correlations, we can determine the subspace, i.e. obtain the error syndrome. [The condition (\ref{rho-4x4}) is actually not needed for distinguishing these subspaces.]

It is important to note that even though within ${\cal Q}_0$ the output signals $I_{1}$ and $I_{2}$ are given by Eq.\ (\ref{I1,2}) with the common term $x_g$, their same-time correlator is $\langle I_{1}(t)\, I_{2}(t+0)\rangle =1$ \cite{Korotkov-2001-spectrum}, and not the naively expected value $x_g^2$. This is because the output noise $\xi_1$ affects the state due to quantum back-action [the first term in Eq.\ (\ref{xg-1})], leading to $\langle \sqrt{\tau_1}\,\xi_1(t)\, x_g(t+0)\rangle=1-x_g^2(t)$ \cite{Korotkov-2001-spectrum}, so that sum of the two terms in the correlator is always $1$. Similarly, all positive correlators (at the same time $t$) are $+1$ and all negative correlators are $-1$. For non-equal times, the correlators $\langle I_k(t_1)\, I_{k'} (t_2)\rangle $ for the corresponding pairs decrease with increasing $|t_1-t_2|$ exponentially with the timescale of the gauge qubit evolution. In particular, in the uniform case (\ref{uniform-4}), from Eqs.\ (\ref{xg-2})--(\ref{zg-2}) and (\ref{I1,2})--(\ref{I3,4}) we find
    \be
    \langle I_1(t_1)\, I_2(t_2)\rangle =  \langle I_3(t_1)\, I_4(t_2)\rangle = \exp (- 2\Gamma_{\rm m} |t_1-t_2|).
    \label{corr-tau}\ee
This formula can be easily derived in the same way as in Ref.\ \cite{Korotkov-2001-spectrum} by noticing from Eqs.\ (\ref{xg-2})--(\ref{zg-2}) that the ensemble-averaged evolution of the gauge qubit is
    \be
    \dot{x}_g = -2\Gamma_{\rm m} x_g, \,\,\,   \dot{y}_g = -4\Gamma_{\rm m} y_g, \,\,\,    \dot{z}_g = -2\Gamma_{\rm m} z_g,
    \ee
so that the $X$-correlator  $\langle I_1(t_1)\, I_2(t_2)\rangle$ and the $Z$-correlator $\langle I_3(t_1)\, I_4(t_2)\rangle$ should both decay in time with the rate $2\Gamma_{\rm m}$.

In the error subspaces the positive cross-correlators have the same value $\exp (- 2\Gamma_{\rm m} |t_1-t_2|)$, while the negative cross-correlators are $-\exp (- 2\Gamma_{\rm m} |t_1-t_2|)$. The cross-correlators for signals measuring orthogonal components of the gauge qubit vanish in all the subspaces, $ \langle I_1(t_1)\, I_3(t_2)\rangle =  \langle I_1(t_1)\, I_4(t_2)\rangle = \langle I_2(t_1)\, I_3(t_2)\rangle =  \langle I_2(t_1)\, I_4(t_2)\rangle =0$.

The correlators in Eq.\ (\ref{corr-tau}) assume ensemble averaging, while we need to monitor the error syndrome in real time from a single realization. The main problem is that the product of noisy outputs is very noisy, so we necessarily need to smoothen out the monitored correlators by time-averaging. For that we used a double-integration  with the bilinear form
    \be
    C_{k\bar{k}} (t)= \iint_{-\infty}^t K(t-t_1, t-t_2)\, I_k(t_1) I_{\bar{k}}(t_2)\, dt_1 dt_2,
    \label{C-general}\ee
where notation $k\bar{k}$ means the channel pairs $12$ or $34$ and the integration kernel $K$ is symmetric. Instead of the general form (\ref{C-general}), it is better to think in terms of integration over the time difference $|t_2-t_1|$ and the mean time $(t_1+t_2)/2$. Obviously, the integral over $|t_2-t_1|$ should be limited to the range $|t_2-t_1|\alt \Gamma_{\rm m}^{-1}$, where the correlator (\ref{corr-tau}) is still significant (so that we do not pick up unnecessary noise).
The integration over $(t_1+t_2)/2$ should be sufficiently long so that the result is not too noisy, but on the other hand a very long integration makes the response time $T_R$ too long.

We have considered two such constructions for the monitored time-integrated correlators,
    \begin{eqnarray}
    && C^{\rm r}_{k\bar{k}}(t)= \frac{1}{T_{\rm c}^{\rm r}} \int^t_{t-T_{\rm c}^{\rm r}} \tilde{C}_{k\bar{k}} (t') \, dt',
    \label{C-r-def}\\
    && C^{\rm e}_{k\bar{k}}(t)= \frac{1}{T_{\rm c}^{\rm e}} \int_{-\infty}^t \tilde{C}_{k{\bar k}} (t')\, e^{-(t-t')/T_{\rm c}^{\rm e}} \, dt',
    \label{C-e-def}\\
    && \tilde{C}_{k\bar{k}}(t)=\frac{1}{2\tau_{\rm c}} \int_{-\infty}^t [I_k(t)I_{\bar{k}}(t')+I_k(t')I_{\bar{k}}(t)]
        \nonumber\\
     && \hspace{2.8cm} \times \, e^{-(t-t')/\tau_{\rm c}} \, dt',
    \label{C-tilde-def}\end{eqnarray}
so that the first (inner) integration (\ref{C-tilde-def}) is always exponential with the time constant $\tau_{\rm c}$ (comparable to $\Gamma_{\rm m}^{-1}$), while the second (outer) integration is either with the rectangular kernel of duration $T_{\rm c}^{\rm r}$ or exponential with the time constant $T_{\rm c}^{\rm e}$ (the time constants $T_{\rm c}^{\rm r,e}$ are at least an order of magnitude longer than $\Gamma_{\rm m}^{-1}$). As will be seen later, the integration with exponential weight (\ref{C-e-def}) provides a better operation of the code than integration with the rectangular weight (\ref{C-r-def}) for typical parameters; however, asymptotically the rectangular integration is better.

From now on, we assume the case without phase back-action and with equal parameters ($\tau_{\rm m}$,  $\Gamma_{\rm m}$) for all four measurement channels, Eq.\ (\ref{uniform-4}), while the quantum efficiency $\eta=(2\Gamma_{\rm m}\tau_{\rm m})^{-1}$ is arbitrary. (It is still rather simple to consider different measurement parameters though formulas become much longer; however, taking into account phase back-action significantly complicates the analysis.) Using Eq.\ (\ref{corr-tau}), it is easy to find average values for $\tilde C_{k\bar{k}}(t)$ in the subspace ${\cal Q}_0$,
    \be
    \langle \tilde C_{12}(t)\rangle =  \langle \tilde C_{34}(t)\rangle =  \langle \tilde C_{k\bar{k}}(t)\rangle = \frac{1}{1+2\Gamma_{\rm m}\tau_{\rm c}}.
    \label{C-tilde-aver}\ee
In the error subspaces this result for cross-correlators is replaced with $\pm 1/(1+2\Gamma_{\rm m}\tau_{\rm c})$, depending on the subspace and pair correlation in the same way as discussed above. The average values for $C_{k\bar{k}}^{\rm r}(t)$ and $C_{k\bar{k}}^{\rm e}(t)$ are the same as for $\tilde{C}_{k\bar{k}}(t)$.

To calculate the noise of $C_{k\bar{k}}^{\rm r,e}(t)$, we first calculate
the time-correlation function for $\tilde C_{k\bar{k}}(t)$. Using the two-time correlators as in Eq.\ (\ref{corr-tau}), four-time correlators \cite{Atalaya-2016-2}
    \be
    \langle I_k(t_1)\, I_{\bar{k}}(t_2)\, I_k(t_3)\, I_{\bar{k}}(t_4)\rangle =
    e^{2\Gamma_{\rm m} (t_1-t_2)} e^{2\Gamma_{\rm m} (t_3-t_4)} \quad
    \ee
for $t_1<t_2<t_3<t_4$, and singularities $\langle I_k(t)\, I_{k}(t')\rangle=\tau_{\rm m}\delta (t-t')$ at  $t\approx t'$, after some algebra we obtain
    \begin{eqnarray}
   && \langle \tilde C_{k\bar{k}} (t_1)\, \tilde C_{k\bar{k}} (t_2) \rangle -\langle  \tilde C_{k\bar{k}}(t) \rangle^2 = e^{-(2\Gamma_{\rm m}+1/\tau_{\rm c})|t_1-t_2|}
      \nonumber \\
      && \hspace{1.5cm} \times \left[ \frac{2\Gamma_{\rm m} \tau_{\rm c}}{(1+2\Gamma_{\rm m} \tau_{\rm c})^2}
       + \frac{\tau_{\rm m}}{4\tau_{\rm c}} +\frac{\tau_{\rm m}}{2\tau_{\rm c}}\, \frac{1}{1+2\Gamma_{\rm m} \tau_{\rm c}}
       \right]
      \nonumber \\
   && \hspace{1.5cm} +\frac{\tau_{\rm m}}{2}\,\delta(t_1-t_2)\left( \frac{\tau_{\rm m}}{2\tau_{\rm c}} +\frac{1}{1+2\Gamma_{\rm m} \tau_{\rm c}} \right). \qquad
    \label{C-tilde-corr}\end{eqnarray}
We see that this correlator decays exponentially with the time constant $(2\Gamma_{\rm m}+1/\tau_{\rm c})^{-1}$. If the second integration in Eqs.\ (\ref{C-r-def}) and (\ref{C-e-def}) is over a much longer period, $T_{\rm c}^{\rm r,e}\gg (2\Gamma_{\rm m})^{-1}$, then the fluctuating part of $\tilde C_{k\bar{k}}(t)$ can be approximately replaced with white noise, which has the same spectral density as the low-frequency spectral density of $\tilde C_{k\bar{k}}(t)$. Therefore, we can use approximation
    \be
    \tilde C_{k\bar{k}}(t) \approx \langle \tilde{C}_{k\bar{k}} \rangle + A \, \tilde{\xi}_{\rm c}(t),
    \label{C-tilde-approx}\ee
where the white noise $\tilde\xi_{\rm c}(t)$ satisfies Eq.\ (\ref{xi-corr}) and
    \begin{eqnarray}
    && A^2 =\int_{-\infty}^{\infty} \left( \langle \tilde C_{k\bar{k}} (0)\, \tilde C_{k\bar{k}} (t) \rangle -\langle  \tilde C_{k\bar{k}} \rangle^2\right) \, dt
    \\
    && \hspace{0.5cm} = \frac{\tau_{\rm m}^2}{4\tau_{\rm c}} + \frac{2\tau_{\rm m}(1+\Gamma_{\rm m} \tau_{\rm c})}{(1+2\Gamma_{\rm m} \tau_{\rm c})^2} +\frac{4\Gamma_{\rm m} \tau_{\rm c}^2}{(1+2\Gamma_{\rm m} \tau_{\rm c})^3} . \qquad
    \label{A2}\end{eqnarray}

This approximation significantly simplifies analysis of noise properties of the monitored integrated correlators $C_{k\bar{k}}^{\rm r,e}(t)$. Note, however, that we neglected possible non-Gaussian contribution to the noise of $\tilde C_{k\bar{k}}$. As will be discussed in Sec.\ \ref{sec:numerics}, numerical simulation shows that the non-Gaussian contribution to the noise slightly changes the obtained below results for the false alarm rate.

Within the approximation (\ref{C-tilde-approx})--(\ref{A2}), we see that independently of the integration kernel of $C_{k\bar{k}}^{\rm r,e}$, we can optimize the signal-to-noise ratio of $\tilde C_{k\bar{k}}$ by minimizing the ratio $A^2/\langle \tilde{C}_{k\bar{k}} \rangle^2$ over $\tau_{\rm c}$. This leads to the following equation for the optimal value $\tau_{\rm c,opt}$:
    \be
    8\eta s^3 (s+2) +4 s^2(1+s)^2 +\eta^{-1}(s^4+2s^3-2s-1)=0,
    \ee
where $s=2\Gamma_{\rm m}\tau_{\rm c,opt}$. Substituting this optimal value into Eqs.\ (\ref{C-tilde-aver}) and (\ref{A2}), we find the optimized $\langle \tilde{C}_{k\bar{k}} \rangle$ and $A$ in Eq.\ (\ref{C-tilde-approx}).

In particular, in the case of ideal detectors, $\eta =1$, the optimal value is $\tau_{\rm c,opt}=0.342/2\Gamma_{\rm m}=0.342\,\tau_{\rm m}$, corresponding to the average signal $\langle \tilde{C}_{k\bar{k}}\rangle=0.745$ and noise power $A^2=2.13\,\tau_{\rm m}$. In the case when $\eta =0.5$, we obtain $\tau_{\rm c,opt}=0.494/2\Gamma_{\rm m}=0.247\,\tau_{\rm m}$, $\langle \tilde{C}_{k\bar{k}}\rangle=0.670$, and $A^2=2.20\,\tau_{\rm m}$.

\subsection{False alarm rate and response time}\label{sec:cont-false-alarm}

Having optimized $\tau_{\rm c}$, let us now discuss the behavior of the monitored integrated cross-correlators  $C_{k\bar{k}}^{\rm r,e}(t)$. Their average values within the subspace ${\cal Q}_0$ do not depend on the integration time  $T_{\rm c}^{\rm r,e}$,
    \be
    \langle C_{k\bar{k}}^{\rm r} \rangle =   \langle C_{k\bar{k}}^{\rm e} \rangle =   \langle \tilde{C}_{k\bar{k}} \rangle,
    \ee
while after a single-qubit error moves the state to an error subspace, the average value for one or both monitored pairs (12 and 34) flips its sign. This error can be detected by observing that the value of a cross-correlator becomes smaller than normal. The most natural criterion for the error detection is crossing of a certain threshold,
    \be
C_{k\bar{k}}^{\rm r,e}(t) < (1-\Theta) \langle \tilde{C}_{k\bar{k}}\rangle.
    \ee
The symmetric threshold corresponds to $\Theta=1$; however, in principle any value within the range $0<\Theta <2$ can be used for the threshold.

Even without the actual error, the monitored correlator  $C_{k\bar{k}}^{\rm r,e}(t)$ can become smaller than the threshold $(1-\Theta) \langle \tilde{C}_{k\bar{k}} \rangle$ due to a big fluctuation. This will be interpreted as an error, and the algorithm will terminate. This will increase the termination rate $\gamma_{\rm term}$ by the rate $\gamma_{\rm f.al.}$ of such ``false alarms'' in each monitored correlator; for example in the model of independent single-qubit errors (Sec.\ \ref{sec:Markovian-cont}) the termination rate will be
        \be
    \gamma_{\rm term} = 2\gamma_{\rm f.al.} + \sum\nolimits_i \left[ \Gamma_i^{(X)}+\Gamma_{i}^{(Y)}+\Gamma_i^{(Z)} \right] ,
    \label{Markovian-term-cont}\ee
where the second term is the rate of actual single-qubit errors and the false alarm rate is doubled because of two monitored correlators with equal and independent noises. Let us now
calculate the false alarm rate $\gamma_{\rm f.al.}$ for one monitored correlator.

It is easy to find the probability distribution $P(C)$ for the correlators $C_{k\bar{k}}^{\rm r,e}(t)$ (within ${\cal Q}_0$) using the white-noise approximation (\ref{C-tilde-approx}),
    \begin{eqnarray}
    && P(C)=\frac{1}{\sqrt{2\pi D^{\rm r,e}_{\rm c}}}\, e^{-(C-\langle C\rangle)^2/2D^{\rm r,e}_{\rm c}},
    \label{P(C)}\\
    && D_{\rm c}^{\rm r} = \frac{A^2}{T_{\rm c}^{\rm r}}, \,\,\, D_{\rm c}^{\rm e} = \frac{A^2}{2T_{\rm c}^{\rm e}},
    \end{eqnarray}
where for brevity we omitted unnecessary subscripts and superscripts and used $\langle C\rangle=\langle\tilde{C}\rangle$. The variance $D_{\rm c}^{\rm r}$ or $D_{\rm c}^{\rm e}$ of the Gaussian distribution has been calculated as the integral of the variances within the shapes (\ref{C-r-def}) or (\ref{C-e-def}). The false alarm rate should be proportional to the probability of being beyond the threshold, $\langle C\rangle -C > \Theta \langle C\rangle$; however, finding the correct prefactor (``attempt frequency'') is not too easy. For that we use the ``first-passage'' approach \cite{Risken-book,Redner-book} and analyze the Fokker-Planck equation for the quasi-stationary first-passage probability distribution $P_{\rm f.p.}(C)$, which has a condition that the threshold has not yet been past and therefore
    \be
    P_{\rm f.p.}[(1-\Theta)\langle C\rangle]=0.
    \label{P-fp-boundary}\ee

The first-passage calculations for exponential integration (\ref{C-e-def}) are relatively easy because the stochastic process $C^{\rm e}(t)$ is Markovian. It can be characterized by the drift velocity $(\langle C\rangle -C^{\rm e})/T_{\rm c}^{\rm e}$ and effective diffusion coefficient $(1/2)(A/T_{\rm c}^{\rm e})^2$. By equating constant probability current (flux) with the first-passage rate $\gamma_{\rm f.al.}$, we write differential equation
    \be
    \frac{\langle C\rangle -C}{T_{\rm c}^{\rm e}} \, P_{\rm f.p.}(C)-\frac{1}{2} (A/T_{\rm c}^{\rm e})^2 \frac{d P_{\rm f.p.}(C)}{d C}=-\gamma_{\rm f.al.}
    \ee
and solve it approximately in the vicinity of the threshold, $C\approx (1-\Theta)\langle C\rangle$, using the boundary condition (\ref{P-fp-boundary}) and also the condition that $P_{\rm f.p.}(C)$ should become practically equal to $P(C)$ from Eq.\ (\ref{P(C)}) away from the threshold. In this way we find the result
    \be
    \gamma_{\rm f.al.}=\frac{\Theta \, \langle C\rangle}{A\sqrt{\pi T_{\rm c}^{\rm e}}}\, e^{-\Theta^2 \langle C\rangle^2 T_{\rm c}^{\rm e}/A^2}
    \label{false-alarm-exp}\ee
for the false alarm rate in the case of exponentially-integrated monitored correlator (\ref{C-e-def}).

The case (\ref{C-r-def}) of the rectangular integration of the correlator is more complicated because the stochastic process $C^{\rm r}(t)$ is not Markovian. However, neglecting non-Markovian effects, for the quasi-stationary distribution $P_{\rm f.p.}(C)$ we can still introduce effective drift velocity $(\langle C\rangle -C^{\rm r})/T_{\rm c}^{\rm r}$ and effective diffusion coefficient $(A/T_{\rm c}^{\rm r})^2$, assuming a big fluctuation, $\langle C\rangle -C^{\rm r} \gg \sqrt{D_{\rm c}^{\rm r}}$. Then in the same way as above we obtain the false alarm rate
        \be
    \gamma_{\rm f.al.}=\frac{\Theta \, \langle C\rangle}{A\sqrt{2\pi T_{\rm c}^{\rm r}}}\, e^{-\Theta^2 \langle C\rangle^2 T_{\rm c}^{\rm r}/2A^2} .
    \label{false-alarm-rect}\ee
Since we had to neglect non-Markovian effects in the derivation, we are not fully sure that the prefactor in Eq.\ (\ref{false-alarm-rect}) is correct; however, numerical simulation confirmed it with the accuracy better than at least 20\%.  Note that using the results of the previous section for the optimization over $\tau_{\rm c}$, for $\eta=1$ we find $\langle C\rangle^2/A^2=0.261\, \tau_{\rm m}^{-1}$ and for $\eta=0.5$ we have $\langle C\rangle^2/A^2=0.203\, \tau_{\rm m}^{-1}$.

In a good quantum error detecting code, the rate $\gamma_{\rm f.al.}$ of false alarms should be less than the rate of actual errors, so that the termination rate (\ref{Markovian-term-cont}) is not significantly increased. (In a quantum error correcting code, $\gamma_{\rm f.al.}$ should be even  less than the rate of logical errors since it contributes to logical errors.) Therefore, the exponent in Eqs.\ (\ref{false-alarm-exp}) and (\ref{false-alarm-rect}) should be rather large, very crudely
    \be
\Theta^2 \langle C\rangle^2 T_{\rm c}^{\rm e}/A^2\simeq \Theta^2 \langle C\rangle^2 T_{\rm c}^{\rm r}/2A^2  \simeq 10{\rm-20}.
    \ee
Increase of the integration time $T_{\rm c}^{\rm e}$ (or $T_{\rm c}^{\rm r}$) decreases the false alarm rate; however, this increases the response time $T_{\rm R}$ (and therefore the rate of logical errors), creating a trade-off between these characteristics of the code operation.

Let us find $T_{\rm R}$ in the simplest way, neglecting the noise. Then, we simply assume that in Eqs.\ (\ref{C-r-def}) and (\ref{C-e-def}) a non-stochastic signal $\tilde{C}(t)$ switches from the constant value $\langle \tilde{C}\rangle$ to $-\langle \tilde{C}\rangle$ at a time moment $t_1$ due to a single-qubit error. By finding the time at which $C^{\rm r,e}(t)$ crosses the threshold $(1-\Theta)\langle \tilde{C}\rangle$ and equating it to $t_1+T_{\rm R}$, we find $T_{\rm R}$. Thus obtained response time for the rectangular and exponential integration is
    \be
    T_{\rm R}^{\rm r}= \frac{\Theta}{2} \, T_{\rm c}^{\rm r}, \,\,\, T_{\rm R}^{\rm e}= T_{\rm c}^{\rm e}\, \ln \frac{2}{2-\Theta} .
    \label{T-R-re}\ee
Since we neglected the noise in finding $T_{\rm R}$, this result becomes inaccurate when $\Theta$ is close to 0 or 2 by $\alt 3 \sqrt{D_{\rm c}^{\rm r,e}}/\langle C\rangle$ [so that the randomness in the distribution (\ref{P(C)}) becomes important].

\subsubsection*{Rectangular vs exponential integration}

Let us compare performance of the rectangular and exponential integrations for the monitored correlator to find out which one is better. As seen from Eqs.\ (\ref{false-alarm-exp}) and (\ref{false-alarm-rect}), the false alarm rates in both cases (for the same threshold) are practically equal if $T_{\rm c}^{\rm r}=2T_{\rm c}^{\rm e}$ (a factor of 2 difference in the prefactor is not very important). Using this relation in Eq.\ (\ref{T-R-re}), we obtain
    \be
     T_{\rm R}^{\rm e} \approx T_{\rm R}^{\rm r}\, \frac{1}{\Theta} \ln \frac{2}{2-\Theta}
    \ee
for the same $\gamma_{\rm f.al.}$
As we see, for the symmetric threshold, $\Theta =1$, the response time for the exponential integration is shorter, $T_{\rm R}^{\rm e} = 0.69\, T_{\rm R}^{\rm r}$. Therefore, exponential integration in the monitored correlator is better than the rectangular integration, providing $31\%$ smaller logical error rate for the same false alarm rate. However, the rectangular integration becomes better than the exponential one for higher thresholds, $\Theta >1.6$; in this case $T_{\rm R}^{\rm r} < T_{\rm R}^{\rm e}$.

Even though the symmetric threshold, $\Theta =1$, seems most natural, the choice of $\Theta$ is rather arbitrary. Let us consider first the integration (\ref{C-e-def}) with exponential kernel and vary $\Theta$, while simultaneously changing the integration timescale $T_{\rm c}^{\rm e}$ to keep the response time $T_{\rm R}^{\rm e}$ constant. Substituting the corresponding value [Eq.\ (\ref{T-R-re})] $T_{\rm c}^{\rm e}=T_{\rm R}^{\rm e}/ \ln [ 2/(2-\Theta)]$ into Eq.\ (\ref{false-alarm-exp}), we find that the false alarm rate $\gamma_{\rm f.al.}$ is proportional  to $\exp \{ -\Theta^2 T_{\rm R}^{\rm e} \langle C\rangle^2 A^{-2}/\ln [2/(2-\Theta)]\}$. Neglecting $\Theta$-dependence in the prefactor, we see that the minimal $\gamma_{\rm f.al.}$ is achieved when $\ln[2/(2-\Theta)]=\Theta/[2(2-\Theta)]$, i.e., at $\Theta_{\rm opt}^{\rm e} =1.43$. Thus, the optimal threshold is not symmetric, and at this optimal $\Theta$ the false alarm rate is
    \be
\gamma_{\rm f.al.}= \frac{0.90 \, \langle C\rangle }{A\sqrt{T_{\rm R}^{\rm e}}}\, e^{-1.63\, T_{\rm R}^{\rm e} \langle C\rangle^2 /A^2} , \,\,\, \Theta_{\rm opt}^{\rm e}=1.43.
    \label{false-alarm-exp-2}\ee
In particular, in the cases $\eta =1$ and $\eta =0.5$ this gives
    \begin{eqnarray}
    &&  \gamma_{\rm f. al} = 0.46 \, (T_{\rm R}^{\rm e}\tau_{\rm m})^{-1/2}\, e^{-0.425 \, T_{\rm R}^{\rm e}/\tau_{\rm m}}, \,\,\, \eta =1, \qquad
     \label{false-alarm-exp-eta1}\\
     && \gamma_{\rm f. al} = 0.41 \, (T_{\rm R}^{\rm e}\tau_{\rm m})^{-1/2}\, e^{-0.331 \, T_{\rm R}^{\rm e}/\tau_{\rm m}}, \,\,\, \eta =0.5. \qquad
       \label{false-alarm-exp-eta05}\end{eqnarray}
As an example, for a desired false alarm rate $\gamma_{\rm f.al.}=10^{-5}\tau_{\rm m}^{-1}$, we need response time $T_{\rm R}^{\rm r}=21.7\, \tau_{\rm m}$ for $\eta=1$ and $T_{\rm R}^{\rm e}=27.2\, \tau_{\rm m}$ for $\eta=0.5$.

However, if the symmetric threshold is chosen, then by using $T_{\rm c}^{\rm e}=T_{\rm R}^{\rm e}/\ln 2$  for the exponential-kernel integration, from Eq.\ (\ref{false-alarm-exp}) we obtain
        \be
    \gamma_{\rm f.al.}=\frac{\langle C\rangle/A}{\sqrt{\pi T_{\rm R}^{\rm e}/\ln 2}}\, e^{-(\langle C\rangle^2/A^2)\, T_{\rm R}^{\rm e}/\ln 2} , \,\,\, \Theta =1.
    \label{false-alarm-exp-3}\ee
For $\eta=1$ and $\eta=0.5$ this gives
    \begin{eqnarray}
&& \gamma_{\rm f.al.}= 0.24 \, (T_{\rm R}^{\rm e}\tau_{\rm m})^{-1/2}  e^{-0.376 \, T_{\rm R}^{\rm e}/\tau_{\rm m}}, \,\,\, \eta =1,
    \\
&& \gamma_{\rm f.al.}= 0.21 \,(T_{\rm R}^{\rm e}\tau_{\rm m})^{-1/2}\, e^{-0.293 \, T_{\rm R}^{\rm e}/\tau_{\rm m}}, \,\,\, \eta=0.5. \qquad
    \end{eqnarray}
Then the desired rate $\gamma_{\rm f.al.}=10^{-5}\tau_{\rm m}^{-1}$ corresponds to the response time $T_{\rm R}^{\rm e}=22.6\, \tau_{\rm m}$ for $\eta=1$ and $T_{\rm R}^{\rm e}=28.3\, \tau_{\rm m}$ for $\eta=0.5$. As we see, the difference in the response time compared with the above case of optimal $\Theta$, is rather minor.

Now let us consider optimization of $\Theta$ in the case of rectangular integration (\ref{C-r-def}). Substituting $T_{\rm c}^{\rm r}$ from Eq.\ (\ref{T-R-re}) into Eq.\ (\ref{false-alarm-rect}), we see that the false alarm rate $\gamma_{\rm f.al.}$ is proportional to $\exp (-\Theta T_{\rm R}^{\rm r} \langle C\rangle ^2/A^2)$, so that it is beneficial to increase $\Theta$ to its maximum possible value of $\Theta =2$. In this case the exponential factor $\exp (-2 T_{\rm R}^{\rm r} \langle C\rangle ^2/A^2)$ is significantly smaller than in Eq.\ (\ref{false-alarm-exp-2})  for the exponential integration for the  same response time. Thus, it seems that for the rectangular integration the optimal threshold is $\Theta =2$, and the performance is better than with the exponential integration.

However, Eq.\ (\ref{T-R-re}) for $T_{\rm R}^{\rm r}$ is significantly inaccurate for $\Theta = 2$. The reason is the fluctuations of $C(t)$, which are on the order of $\pm \sqrt{D_{\rm c}^{\rm r}} = \pm A/\sqrt{T_{\rm c}^{\rm r}}$ [see Eq.\ (\ref{P(C)})]. For the negative fluctuation, the crossing of the threshold occurs earlier by $\sim \sqrt{D_{\rm c}^{\rm r}}/(2\langle C\rangle /T_{\rm c}^{\rm r})$, while for the positive fluctuation the crossing occurs later by $\sim T_{\rm c}^{\rm r}$, which is much longer (crudely by the factor $\sqrt{T_{\rm c}^{\rm r}/\tau_{\rm m}}$). This asymmetry significantly increases the average response time $T_{\rm R}^{\rm r}$.

To avoid this problem, let us shift the threshold by 2 standard deviations (so that we can neglect the fluctuations), then $\Theta = 2 - 2\sqrt{D_{\rm c}^{\rm r}}/\langle C \rangle$. In this case from Eqs.\ (\ref{false-alarm-rect}) and (\ref{T-R-re}) we obtain approximately
    \be
    \gamma_{\rm f. al.} \approx \frac{\sqrt{2}\,\langle C\rangle}{A\sqrt{\pi T_{\rm R}^{\rm r}}} \, \exp \left (-\frac{2\langle C\rangle^2 T_{\rm R}^{\rm r}}{A^2} + \frac{2\langle C\rangle \sqrt{T_{\rm R}^{\rm r}}}{A} -1 \right) .
    \label{false-alarm-rect-2}\ee
For the desired false alarm rate $\gamma_{\rm f.al.}=10^{-5}\tau_{\rm m}^{-1}$, this gives the response time $T_{\rm R}^{\rm r}=25.1\, \tau_{\rm m}$ for $\eta=1$ and $T_{\rm R}^{\rm e}=31.6\, \tau_{\rm m}$ for $\eta=0.5$. Somewhat surprisingly, this response time is longer than even for the exponential integration with symmetric threshold, Eq.\ (\ref{false-alarm-exp-3}), in spite of faster decaying main exponential term in Eq.\ (\ref{false-alarm-rect-2}). The reason is that our shift of the threshold by two standard deviations is quite significant for these parameters, leading to a significant positive term within the exponent of Eq.\ (\ref{false-alarm-rect-2}).

Therefore, even though asymptotically the rectangular integration (\ref{C-r-def}) for the monitored correlator (with the threshold $\Theta$ approaching 2) is better than the exponential integration (\ref{C-e-def}), for our typical parameters the exponential integration is better. Moreover, since for the exponential integration there is no big difference between the results for the optimal $\Theta$ [Eq.\ (\ref{false-alarm-exp-2})] and for the symmetric threshold [Eq.\ (\ref{false-alarm-exp-3})], and since choosing the symmetric threshold avoids possible problems with $\Theta$ being too close to 2 (as for the rectangular integration case), we conclude that the symmetric threshold, $\Theta =1$,  is a good choice.

Note that besides the definition (\ref{C-tilde-def}) for the signal $\tilde C_{k\bar{k}}(t)$ (which is then integrated to give the monitored correlators), we also considered the definition
    \be
    \tilde{C}_{k\bar{k}}(t)= \tilde{I}_k(t)\,\tilde{I}_{\bar{k}}(t), \,\, \tilde{I}_k(t)= \frac{1}{\tau_{\rm c}} \int_{-\infty}^t I_k(t')\,e^{-(t-t')/\tau_{\rm c}} dt',
    \label{C-tilde-def-2}\ee
which still leads to a bilinear form (\ref{C-general}) after applying integration (\ref{C-r-def}) or (\ref{C-e-def}). The definition (\ref{C-tilde-def-2}) is more natural for an experimental realization. It can also be naturally generalized to the nine-qubit Bacon-Shor code with continuous measurement, which will be able to operate as a quantum error correcting code (not only detecting). Even though Eq.\ (\ref{C-tilde-def-2}) formally contains two integrations in contrast to the single integration in Eq.\ (\ref{C-tilde-def}), the important integration is only over the time difference between the two channels, while the integration over the running time is anyway repeated in forming $C_{k\bar{k}}(t)$. As a result, the integrated correlator  $C_{k\bar{k}}(t)$ in Eq.\ (\ref{C-r-def}) is practically the same when either Eq.\ (\ref{C-tilde-def}) or (\ref{C-tilde-def-2}) is used for $\tilde{C}_{k\bar{k}}$ if $T_{\rm c}^{\rm r}\gg \tau_{\rm c}$ (the difference is only near the edges of the integration, with the relative difference on the order of $\tau_{\rm c}/T_{\rm c}^{\rm r}$). For the exponential integration in Eq.\ (\ref{C-e-def}), the relative difference for $C_{k\bar{k}}(t)$ is similarly on the order of $\tau_{\rm c}/T_{\rm c}^{\rm e}$. Therefore, we can still use Eqs.\ (\ref{C-tilde-aver}) and (\ref{A2}) for the signal and low-frequency noise of $\tilde{C}_{k\bar{k}}(t)$ defined via Eq.\ (\ref{C-tilde-def-2}), and thus all results derived in this section remain (approximately) valid.

\subsection{Monte Carlo simulation results}\label{sec:numerics}

To check the developed above (approximate) theory for the termination and logical error rates, we have performed quantum trajectory simulations for the full density matrix $\rho(t)$ of the four-qubit system. For each time step $\delta t$, the density matrix $\rho(t+\delta t)$ is obtained from $\rho(t)$ by consecutively applying the random quantum Bayesian updates, corresponding to measurements of the gauge operators $G_k$ ~\cite{Korotkov-2002ent}. Then, to the resulting density matrix we apply an extra evolution to account for the environmental decoherence within the same timestep; for that we use the Lindblad equation [see Eqs.~\eqref{Lindblad-1}--\eqref{Lindblad-2}],  obtaining $\rho(t+\delta t)$ up to second order in $\delta t$. In the simulations we use the orthonormal basis, introduced in section~\ref{sec:system}, neglect the phase back-action, and assume ideal measurements of equal strength $\Gamma_{\rm m}=1/2\tau_{\rm m}$ for all gauge operators. The time step is $\delta t = 5\cdot10^{-3}\, \Gamma_{\rm m}^{-1}$.

The time-integrated correlators $C_{12}(t)$ and $C_{34}(t)$ are computed using Eqs.~\eqref{C-r-def}--\eqref{C-tilde-def} for each of $10^4$--$10^5$ trajectories. For a given duration $T$ of the process, ``good'' (no-detected-error) trajectories are selected by the condition that the correlators for both channels are above the threshold $(1-\Theta)\langle C\rangle$ for the whole duration $T$. The results presented below are for the symmetric case, $\Theta =1$.
The relative number of no-detected-error trajectories gives (approximately) the success probability $P_{\rm success}(T)$; by fitting this numerical dependence to the exponential decay of Eq.~\eqref{success-prob}, we obtain the termination rate $\gamma_{\rm term}$.
In particular, in the absence of decoherence, $\gamma_{\rm term}$ is twice  the false alarm rate $\gamma_{\rm f.al.}$ per channel.

\begin{figure}[tb]
\includegraphics[width=\linewidth]{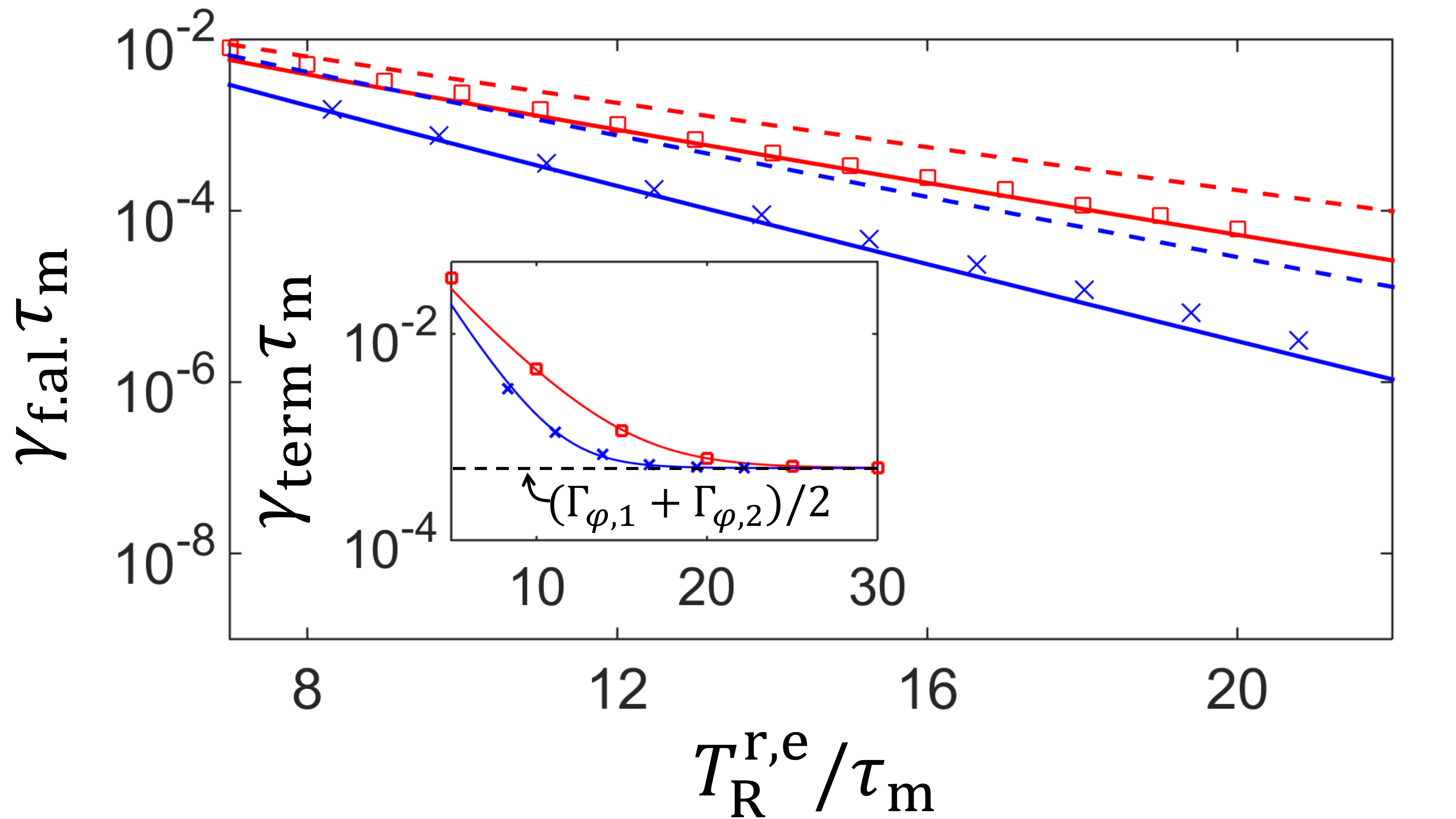}
\caption{(Color online) The false alarm rate $\gamma_{\rm f.al.}$ as a function of the response time $T_{\rm R}^{\rm r,e}$ (normalization involves the collapse time $\tau_{\rm m}$). Numerical results for the rectangular and  exponential correlator integrations are shown by red squares and blue crosses, respectively. The (upper) red and (lower) blue dashed lines represent the analytics, Eqs.\ \eqref{false-alarm-rect} and  \eqref{false-alarm-exp}. The solid lines include correction factors in the analytical formulas (see main text). The inset shows the termination rate $\gamma_{\rm term}$ as a function of $T_{\rm R}^{\rm r, e}$ (the same horizontal axis as in the main panel) for the case when qubits 1 and 2 are subject to pure dephasing with $\Gamma_{\varphi,1} = \Gamma_{\varphi,2}=10^{-3}\Gamma_{\rm m}$ (no decoherence in the main panel); the horizontal dashed line corresponds to single-qubit errors, $\gamma_{\rm term}= (\Gamma_{\varphi,1} + \Gamma_{\varphi,2})/2$. We used $\Theta=1$, $\eta=1$ and $\tau_{\rm c} = \tau_{\rm c, opt}$.
}
\label{fig:gamma_fa_term}
\end{figure}

Figure \ref{fig:gamma_fa_term} shows thus calculated false alarm rate $\gamma_{\rm f.al.}$ as a function of the response time $T_{\rm R}^{\rm r,e}$ [obtained from the actual integration time $T_{\rm c}^{\rm r,e}$ via Eq.\ (\ref{T-R-re})] for monitoring the time-integrated correlators with rectangular (red squares) and exponential (blue crosses) kernels.
The red and blue dashed lines show the analytical formulas \eqref{false-alarm-rect} and \eqref{false-alarm-exp}, respectively.
The numerical results indicate that the analytical formulas slightly underestimate the coefficients in the exponents. We have found that this discrepancy between numerics and analytics is due to non-Gaussian fluctuations of $C^{\rm r,e}_{k\bar{k}}(t)$ [which were assumed to be Gaussian in the analytical derivation because of the approximation (\ref{C-tilde-approx}) for $\tilde{C}_{k\bar{k}}(t)$]. In particular, for $\tau_{\rm c}=\tau_{\rm c,opt}$ and $\eta =1$, we numerically calculated the third cumulant $\kappa_3=\langle C^3\rangle-3\langle C^2\rangle \langle C\rangle +2\langle C\rangle^2$ for $C^{\rm r}_{k\bar k}(t)$ and $C^{\rm e}_{k\bar k}(t)$, obtaining $\kappa_3 \approx 1.05/(\Gamma_{\rm m}T^{\rm r}_{\rm c})^2$ and $\kappa_3 \approx 0.34/(\Gamma_{\rm m}T_{\rm c}^{\rm e})^2$  for the cases of rectangular and exponential integrations, respectively ($\kappa_3=0$ for a Gaussian process). The non-zero third cumulant leads to the correction factor $1 +  \langle C\rangle\kappa_3/3\kappa_2^2$ in the exponent for $\gamma_{\rm f.al.}$ in Eqs.\ \eqref{false-alarm-rect} and \eqref{false-alarm-exp}, where the second cumulant is $\kappa_2=A^2/T_{\rm c}^{\rm r}$ and $\kappa_2=A^2/2T_{\rm c}^{\rm e}$ for these two cases. This gives the correction factors of 1.23 and 1.30 to the exponents of Eqs.\ \eqref{false-alarm-rect} and \eqref{false-alarm-exp}, respectively (for $\tau_{\rm c}=\tau_{\rm c,opt}$ and $\eta =1$). The red and blue solid lines in Fig.\ \ref{fig:gamma_fa_term} show the analytical results with account of these corrections, which agree well with the numerical results. Note that the main figure shows the false alarm rate $\gamma_{\rm f.al.}$ calculated in the absence of decoherence, while the inset shows the termination rate in the presence of dephasing in qubits 1 and 2 with the rates $\Gamma_{\varphi,1}=\Gamma_{\varphi,2}=10^{-3\,}\Gamma_{\rm m}$. In this case, for small response times the termination rate is dominated by false alarms, but for large response times the termination rate converges to the rate of single-qubit errors (horizontal dashed line) -- see Eq.~\eqref{Markovian-term-cont}.

The logical error rates have been calculated numerically in the following way. First,
to extract the logical qubit state (for no-detected-error trajectories) from the four-qubit density matrix $\rho$ at each time $T$, we apply the transformation $\rho(T)\to\tilde\rho(T) = \Pi_{++}\rho(T)\Pi_{++} + \Pi_{--}\rho(T)\Pi_{--}$, where $\Pi_{++} = (\openone + G_3)(\openone + G_4)/4$ and $\Pi_{--} = (\openone - G_3)(\openone - G_4)/4$ are projection operators. This transformation corresponds to applying projective measurements of $G_3$ and $G_4$ at time $T$ in the decoding procedure and selecting only outcomes with the same results. Then the
logical qubit state is extracted from the $4\times 4$ block of $\tilde\rho(T)$,  corresponding to the code space $\mathcal{Q}_0$, by tracing out the gauge qubit. The resulting Bloch coordinates of the logical qubit are given by the equations
\begin{align}
x_{\rm L} &= 2\,{\rm Re} \frac{\left\langle \tilde{\rho}_{13} + \tilde{\rho}_{24}\right\rangle}{\left\langle{\rm Tr}\, \tilde{\rho}\right\rangle},
    \,\,\,
y_{\rm L} =-2\,{\rm Im} \frac{\left\langle \tilde{\rho}_{13} + \tilde{\rho}_{24}\right\rangle}{\left\langle{\rm Tr}\, \tilde{\rho}\right\rangle},
    \label{logical-qubit-state}\\
z_{\rm L} &=\frac{\left\langle\tilde{\rho}_{11} + \tilde{\rho}_{22} - \tilde{\rho}_{33} - \tilde{\rho}_{44}\right\rangle}{\left\langle{\rm Tr} \, \tilde{\rho}\right\rangle},
    \end{align}
where the indices correspond to the basis (\ref{phi-1})--(\ref{phi-4}) and averaging is over trajectories with no detected errors. From $\{x_{\rm L}(T),\, y_{\rm L}(T),\, z_{\rm L}(T)\}$ for four initial logical states, we calculate the quantum process matrix $\chi (T)$ for the logical qubit state evolution~\cite{N-C-book,Kofman-2009}. Then the logical error rates $\gamma_X$, $\gamma_Y$, and $\gamma_Z$ are extracted from the linear dependence on time $T$ of the diagonal elements $\chi_{XX}$, $\chi_{YY}$, and $\chi_{ZZ}$. Note that we normalize the process matrix, $\chi \rightarrow \chi /{\rm Tr} (\chi )$ after checking that the success probability does not depend on the initial logical state.

\begin{figure}[tb]
\includegraphics[width=\linewidth, trim=0.5cm 0.0cm 0.0cm 0.0cm,clip=true]{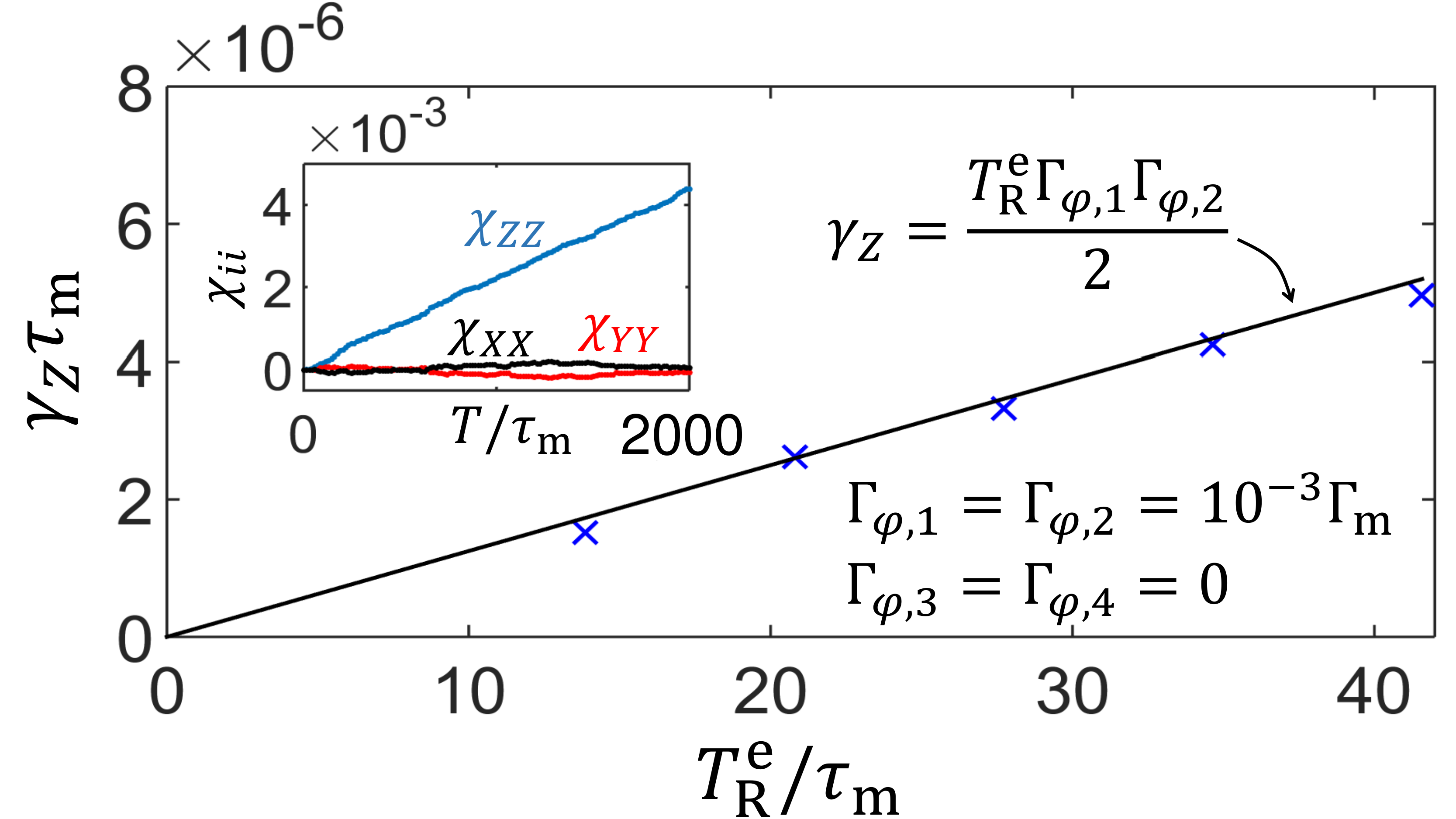}
\caption{(Color online) The rate $\gamma_{Z}$ of the logical $Z$-error as a function of the response time $T_{\rm R}^{\rm e}$ for the case when qubits 1 and 2 are subject to pure dephasing with $\Gamma_{\varphi ,1}=\Gamma_{\varphi ,2}=10^{-3}\Gamma_{\rm m}$. The crosses show numerical results, the solid line is the analytical result, Eq.~\eqref{gamma-Z-dephasing}. The inset shows the diagonal elements of the process matrix ($\chi_{XX}$, $\chi_{YY}$, and $\chi_{ZZ}$) as functions of time $T$ for $T_{\rm R}^{\rm e} = 20.8\tau_{\rm m}$. The rate $\gamma_Z$ is calculated from the slope of  $\chi_{ZZ}(T)$. We used $\Theta =1$, $\eta=1$, $\tau_{\rm c}=\tau_{\rm c, opt}$, and the exponential kernel for the correlators. }
\label{fig:gamma_Z_deph}
\end{figure}

We have checked our analytical formulas for the logical error rates (Sec.\ \ref{sec:cont-logical-errors}) against the numerical results for the cases of pure dephasing and energy relaxation (amplitude damping). For pure dephasing, we have found that Eqs.\ \eqref{gamma-XY-dephasing}--\eqref{gamma-Z-dephasing} agree well with the numerical results for the logical error rates. As an example, Fig.\ \ref{fig:gamma_Z_deph} shows dependence of the logical $Z$-error rate $\gamma_{\rm Z}$ on the response time $T_{\rm R}^{\rm e}$ [obtained from the exponential integration time $T_{\rm c}^{\rm e}$ via Eq.\ (\ref{T-R-re})] for the case when the qubits 1 and 2 are subject to dephasing with $\Gamma_{\varphi ,1}=\Gamma_{\varphi , 2}=10^{-3}\Gamma_{\rm m}$. We see that the agreement between the numerics (crosses) and analytics (line) is quite good. The inset in Fig.~\ref{fig:gamma_Z_deph} shows time-dependence of all three diagonal elements of the process matrix $\chi$. We see that even though numerical values of $\chi_{XX}$ and $\chi_{YY}$ are not exactly zero (as they should be analytically), they do not grow with the time $T$; their non-zero values are due to statistical noise in the Monte Carlo simulations. The numerical off-diagonal elements of the $\chi$-matrix are zero (not shown).
Similar results are obtained when other pairs of qubits are subject to pure dephasing.

\begin{figure}[!tb]
\begin{tabular}{cc}
\includegraphics[width=\linewidth, trim=0.cm 0.0cm 0.0cm 0.0cm,clip=true]{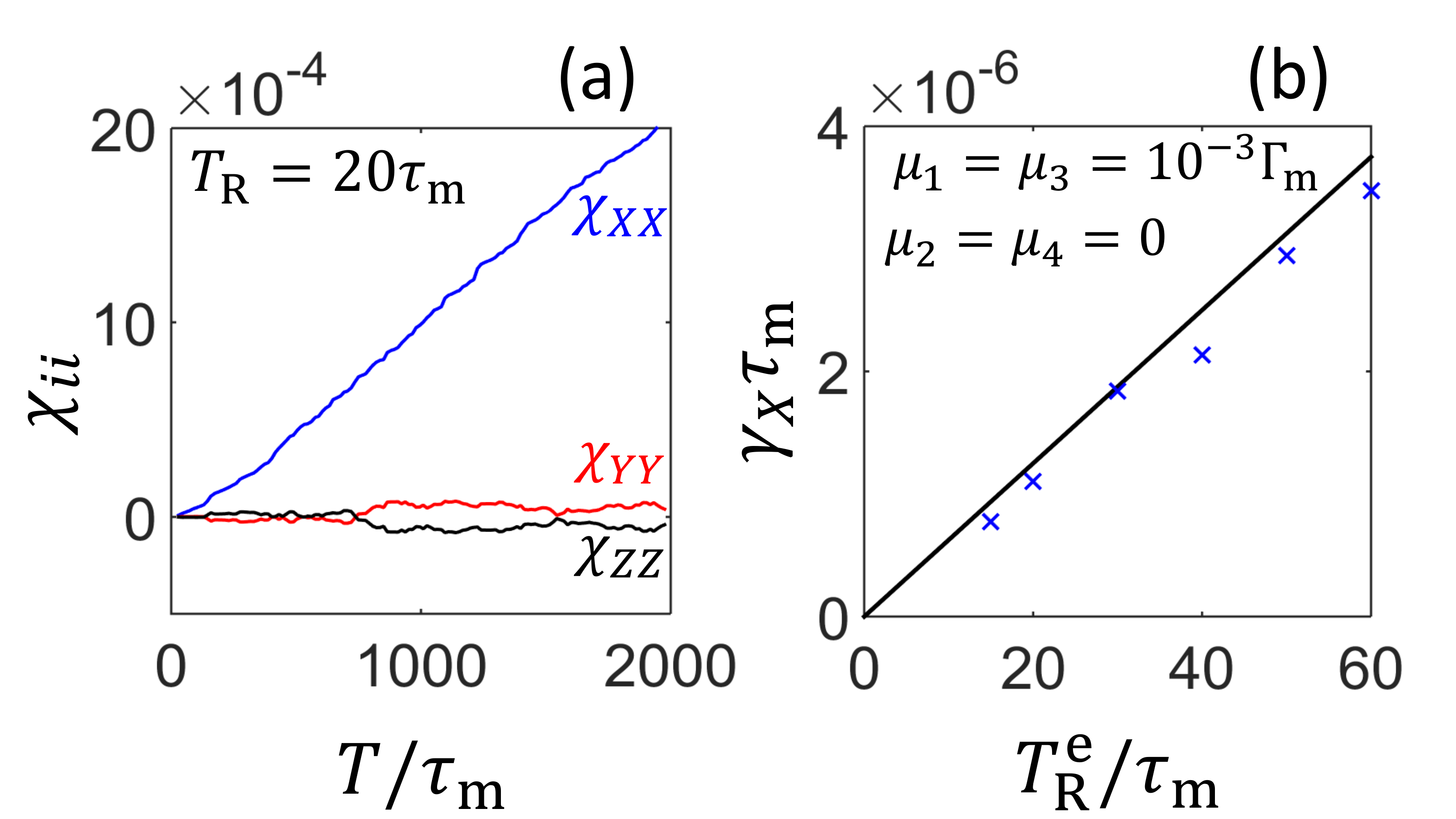}
\\
\includegraphics[width=\linewidth, trim=0.cm 0.0cm 0.0cm 0.5cm,clip=true]{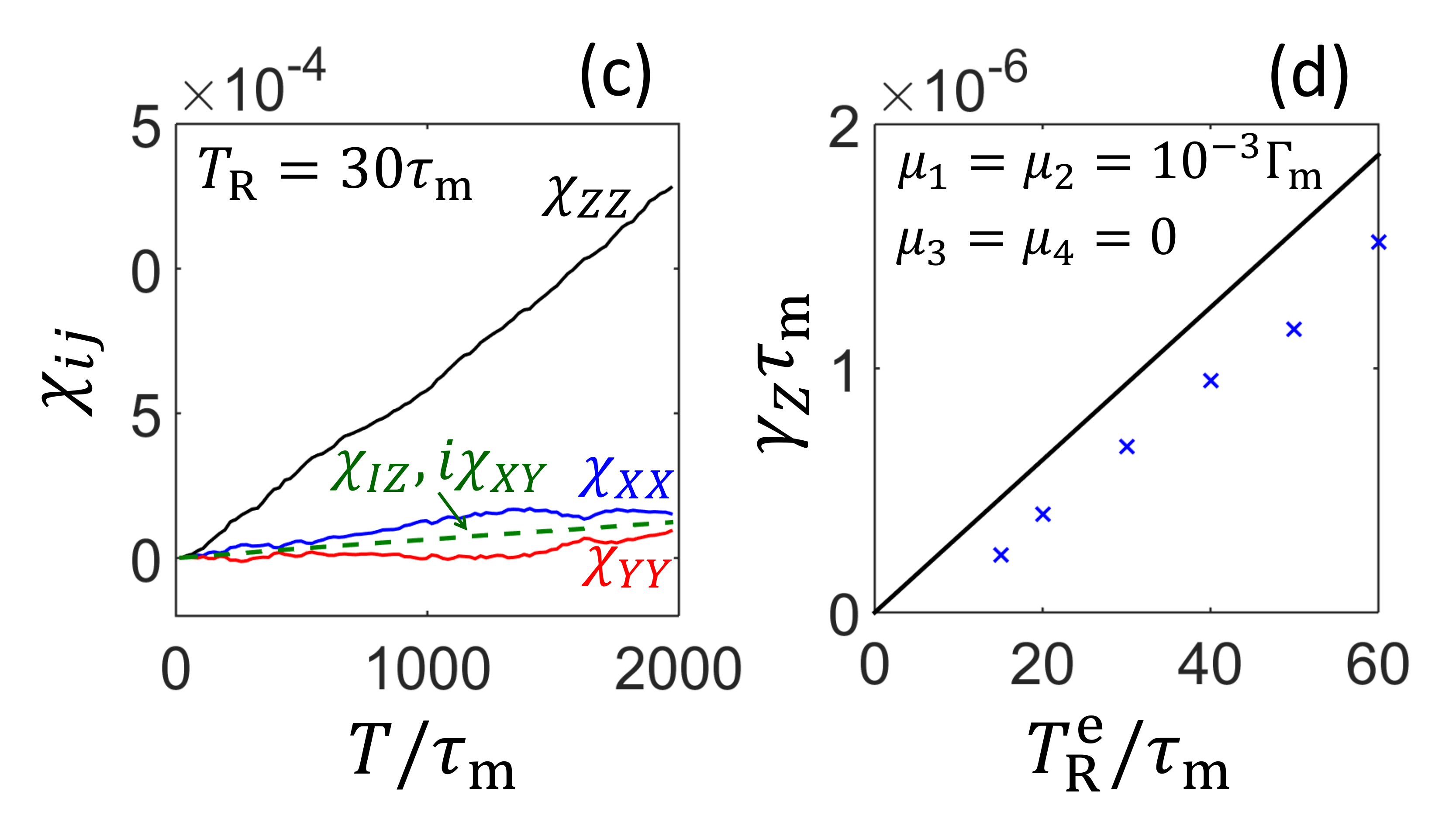}
\end{tabular}
\caption{(Color online) The $\chi$-matrix elements and logical error rate for the case of qubit energy relaxation (amplitude damping). In panels (a) and (b) the energy relaxation affects only qubits 1 and 3, $\mu_1=\mu_3=10^{-3}\Gamma_{\rm m}$; in panels (c) and (d) it affects only qubits 1 and 2, $\mu_1=\mu_2=10^{-3}\Gamma_{\rm m}$.
Panels (a) and  (c) show numerically calculated components of the $\chi$-matrix as functions of time $T$; the diagonal elements are depicted by solid lines. The only non-zero off-diagonal elements are $\chi_{IZ}$ (real) and $\chi_{XY}$ (imaginary) in panel (c), shown by (almost coinciding) dashed green lines. Panels (b) and (d) show, respectively, the logical error rates $\gamma_X$ and $\gamma_Z$ as functions of the response time $T_{\rm R}^{\rm e}$. The crosses represent numerical results and the solid lines represent analytical formulas, Eqs.~\eqref{relax-X-cont} and \eqref{relax-Z-cont}. We used $\Theta =1$, $\eta=1$, $\tau_{\rm c}=\tau_{\rm c, opt}$, and the exponential kernel for the correlators.
 }
\label{fig:gamma_Z_energy_relax}
\end{figure}

Numerical results for the case of energy relaxation in physical qubits also agree with analytical results (\ref{relax-X-cont})--(\ref{relax-Z-cont}); however, there are minor deviations discussed below. Let us first assume that only qubits 1 and 3 are subject to energy relaxation. For this case we expect only logical $X$-errors -- see Eqs.\ (\ref{relax-X-cont})--(\ref{relax-Z-cont}). Indeed, our numerical results shown in Fig.\ \ref{fig:gamma_Z_energy_relax}(a) indicate that out of the diagonal elements of $\chi (T)$, only $\chi_{XX}$ exhibits linear scaling with time $T$ (non-zero values of $\chi_{YY}$ and $\chi_{ZZ}$ are due to inaccuracy of Monte Carlo simulations), and the off-diagonal elements are zero (not shown). The extracted logical error rate $\gamma_X$ is shown by crosses in Fig.~\ref{fig:gamma_Z_energy_relax}(b) as a function of the response time $T_{\rm R}^{\rm e}$ (the exponential integration of the correlators is used for all panels of Fig.\ \ref{fig:gamma_Z_energy_relax}). The agreement with the analytical formula \eqref{relax-X-cont} [solid line in Fig.\ \ref{fig:gamma_Z_energy_relax}(b)] is good. However, the agreement is not so good for the case of energy relaxation in the qubits 1 and 2, presented in Figs.\ \ref{fig:gamma_Z_energy_relax}(c) and  \ref{fig:gamma_Z_energy_relax}(d).
Figure \ref{fig:gamma_Z_energy_relax}(c) indicates that even though elements $\chi_{XX}$ and $\chi_{YY}$ are much smaller than the main element $\chi_{ZZ}$, they still increase with time, in contrast to what is expected from Eqs.\ (\ref{relax-X-cont})--(\ref{relax-Z-cont}). We have also found several small but non-zero off-diagonal elements, linearly increasing with time $T$; numerical results can be fitted well by formulas $\chi_{IZ}(T) =\chi_{ZI}(T) = T\mu_1\mu_2/8\Gamma_{\rm m}$ and $\chi_{XY}(T) = -\chi_{YX}(T)= -iT\mu_1\mu_2/8\Gamma_{\rm m}$ (other off-diagonal elements are practically zero). Note that these small elements are not proportional to the response time, in contrast to the main element $\chi_{ZZ}=TT_{\rm R}^{\rm e}\mu_1\mu_2/8$. Figure \ref{fig:gamma_Z_energy_relax}(d) shows the numerical logical error rate $\gamma_Z$ (crosses) extracted from the linear dependence $\chi_{ZZ}(T)$. The analytical result given by Eq.\ (\ref{relax-Z-cont}) is shown by the solid line. There is apparently a shift between the numerical and analytical results. We do not know what is exactly the reason for this discrepancy. For example, it can be because an error can be detected due to correlator noise even after the second single-qubit error occurred. It can also be related to no-jump evolution, which was neglected in the analytical derivation in Sec.\ \ref{sec:relax-cont}, which included only the effects scaling linearly with $T_{\rm R}$. The numerical results for the energy relaxation in qubit 1 and 4 are similar to the results presented in Fig.\  \ref{fig:gamma_Z_energy_relax}, with dependence $\gamma_X (T_{\rm R}^{\rm e})$ agreeing well with analytics similar to Fig.\ \ref{fig:gamma_Z_energy_relax}(b) and  $\gamma_Y (T_{\rm R}^{\rm e})$ showing a shift from analytics similar to Fig.\ \ref{fig:gamma_Z_energy_relax}(d).
In spite of the minor deviations, we conclude that numerical results agree with the (approximate) analytics (\ref{relax-X-cont})--(\ref{relax-Z-cont}).

Note that our analytical derivation is based on the picture of abrupt jumps between the code space and error subspaces because of single-qubit errors, while in the numerical simulations we use the Lindblad equation to describe continuous evolution due to decoherence. Nevertheless, in the simulations we clearly see almost abrupt state transitions between the subspaces, which are caused by the interplay between the decoherence, which mixes the subspaces, and measurement, which gradually collapses the state into only one subspace. The timescale of the transitions is much shorter than the response time needed for correlators to report the transition.

\subsection{Comparison with projective measurement case}\label{sec:cont-comparison}

For a quantum error detecting code, there are two main characteristics of performance: success probability (probability that no errors have been detected) and probability of a logical error (assuming that no errors have been detected). Since in our case the probability of a detected error and probability of a logical error both linearly depend on time (for a sufficiently short time), it is more convenient to use the termination rate [see Eq.\ (\ref{success-prob})] and the logical error rate.

To compare operations of the 4-qubit Bacon-Shor code with projective and continuous measurements, let us use the model of uncorrelated Markovian errors. The rates of logical $X$, $Y$, and $Z$ errors for the projective-measurement case are given by Eqs. (\ref{proj-X-Mark})--(\ref{proj-Z-Mark}), while for the continuous-measurement case they are given by Eqs.\ (\ref{cont-X-Mark})--(\ref{cont-Z-Mark}). In general, the formulas in the two cases are similar to each other, with the projective-measurement half-cycle time $\Delta t$ (or $\Delta t/2$) replaced  with the response time $T_{\rm R}$ for the continuous measurement. (Some difference in the formulas is because in the continuous-measurement mode, the response time $T_{\rm R}$ is the same for any single-qubit error, while in the projective-measurement mode, $Y_i$-errors are detected on average twice sooner than $X_i$ or $Z_i$ errors.)
Since the formulas are slightly different, let us assume equal rates for errors of all types in all qubits (depolarizing channel), as in Eqs.\ (\ref{gamma-L-depol-proj}) and (\ref{gamma-L-depol-cont}).  Then we see that the ratio of the total logical error rates $\gamma_{\rm L}$ for the projective and continuous measurements is $\gamma_{\rm L, cont}/\gamma_{\rm L, proj}=(14/11)\, T_{\rm R}/\Delta t$. In particular, for the continuous-measurement correlator integration (\ref{C-e-def}) with exponential weight and symmetric threshold, $\Theta=1$, this ratio of the logical error rates is
    \be
    \frac{\gamma_{\rm L,cont}}{\gamma_{\rm L,proj}}=  \frac{14\, \ln 2}{11} \, \frac{T_{\rm c}^{\rm e}}{\Delta t} \approx 0.9\,  \frac{T_{\rm c}^{\rm e}}{\Delta t},
    \label{log-err-ratio}\ee
where $T_{\rm c}^{\rm e}$ is the correlator integration time.

Besides the logical error rates $\gamma_{\rm L}$, we need to compare the termination rates $\gamma_{\rm term}$. For the projective-measurement case, $\gamma_{\rm term}$ is (almost) the sum of single-qubit rates, so for the depolarizing channel with $\Gamma_i^{(X)}=\Gamma_i^{(Y)}=\Gamma_i^{(Z)}=\Gamma_{\rm d}/3$ it is $\gamma_{\rm term}=4\Gamma_{\rm d}$. In the continuous-measurement case, $\gamma_{\rm term}$ is increased by the false alarm rate for each of two monitored correlators, so using Eq. (\ref{false-alarm-exp}) for the exponential integration of the correlator with $\Theta =1$, we obtain the ratio
    \be
     \frac{\gamma_{\rm term,cont}}{\gamma_{\rm term,proj}}= 1+2\,\frac{\langle C\rangle/A}{4\Gamma_{\rm d}\sqrt{\pi T_{\rm c}^{\rm e}}} \, e^{-T_{\rm c}^{\rm e}\langle C\rangle^2 /A^2} ,
    \label{term-rate-ratio}\ee
where, as discussed above, $\langle C\rangle^2 /A^2=0.26\, \tau_{\rm m}^{-1}$ for ideal detectors, $\eta =1$, and $\langle C\rangle^2 /A^2=0.20\, \tau_{\rm m}^{-1}$ for detectors with efficiency $\eta =0.5$. Note that as discussed in the previous section, the exponential suppression of the false alarm rate is actually about 30\% stronger due to non-Gaussian effects (which improves the operation); however, for simplicity we neglect this correction here.

\begin{figure}[tb]
\includegraphics[width=8.5cm, trim=0.5cm 0.0cm 0.0cm 0.0cm,clip=true]{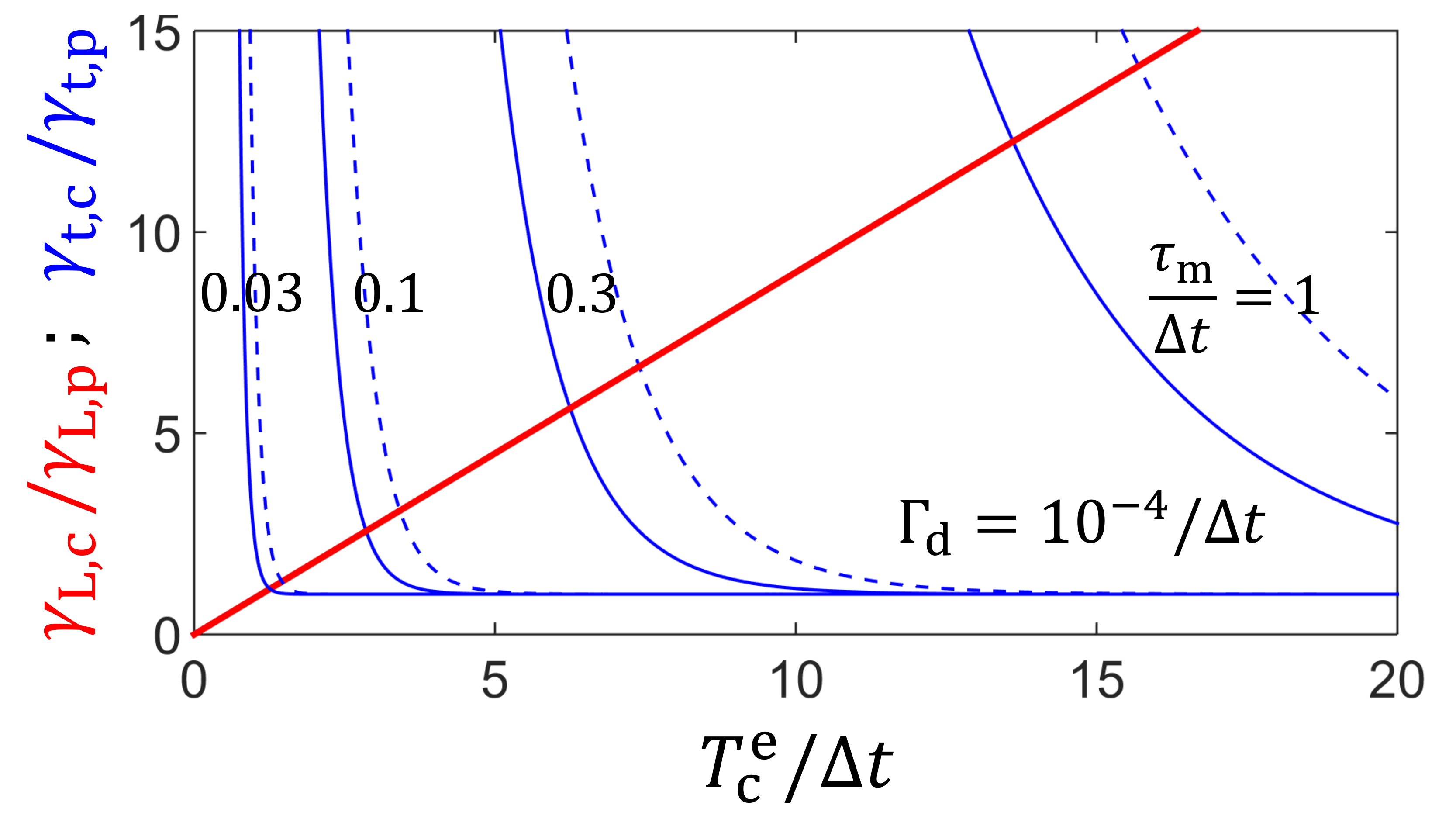}
\caption{(Color online) Trade-off between the logical error rate and termination rate for the Bacon-Shor code with continuous measurements.  Thick red line: ratio $\gamma_{\rm L, cont}/\gamma_{\rm L, proj}$ of the logical error rates for continuous and projective measurements [Eq.\ (\ref{log-err-ratio})], as a function of the correlator integration time $T_{\rm c}^{\rm e}$ normalized by the projective-measurement half-cycle time $\Delta t$. Thin blue lines: ratio $\gamma_{\rm term, cont}/\gamma_{\rm term, proj}$  of the termination rates for the continuous and projective measurements [Eq.\ (\ref{term-rate-ratio})] for several values of the collapse (``measurement'') time $\tau_{\rm m}$: $\tau_{\rm m}/\Delta t=1$, 0.3, 0.1, and 0.03 (from right to left) and quantum efficiency $\eta=1$ (solid lines) or $\eta=0.5$ (dashed lines). We assume $\Gamma_{\rm d}=10^{-4}/\Delta t$ and $\Theta=1$.
}
\label{fig:term-logical}
\end{figure}

Figure \ref{fig:term-logical} shows the ratios of the logical error and termination rates, Eqs.\ (\ref{log-err-ratio}) and (\ref{term-rate-ratio}), as functions of the correlator integration time $T_{\rm c}^{\rm e}$ for several values of the collapse (``measurement'')  time $\tau_{\rm m}$ for each detector, assuming $\Gamma_{\rm d}=10^{-4}/\Delta t$ and $\eta=1$ (thin solid lines) or $\eta=0.5$ (thin dashed lines). We see that if $\tau_{\rm m}=\Delta t$, then in order to keep $\gamma_{\rm term,cont}/\gamma_{\rm term,proj} \alt 3$, we need to choose $T_{\rm c}^{\rm e}/\Delta t \agt 20$, and correspondingly the logical error rate is also a factor of 20 larger than in the projective-measurement case. However, if $\tau_{\rm m}=0.03\, \Delta t$, then $\gamma_{\rm term,cont}/\gamma_{\rm term,proj} \sim 3$ corresponds to $\gamma_{\rm L,cont}/\gamma_{\rm L, proj}\sim 1$.

We see that for comparable operations of the code in the continuous and projective measurement cases, we need a quite strong continuous measurement, $\tau_{\rm m}\sim \Delta t/30$ (the non-Gaussian corrections increase this estimate to $\sim \Delta t/20$). Even though this may  seem as a disadvantage of using continuous measurement, actually the same problem is hidden in the  assumption of an instantaneous projective measurement. Since any measurement in circuit QED architecture for superconducting qubits is physically continuous, for a ``projective'' measurement with infidelity of $\sim 10^{-5}$ we need duration $\sim 5\tau_{\rm m}$. In the conventional code with projective measurements, this duration is assumed to be much shorter than the half-cycle $\Delta t$. Therefore, our result of $\Delta t \sim 20\,\tau_{\rm m}$ is not surprising, and the same or larger ratio is implicitly assumed in the conventional code with projective measurements. (Note that the logical qubit is not protected during experimental ``projective'' measurement of two-qubit operators, which includes quantum gates between the code qubits and ancillary qubits.)

\section{Conclusion}\label{sec:Conclusion}

In this paper we have analyzed the operation of a four-qubit Bacon-Shor code, in which projective measurements of two-qubit operators are replaced with their continuous measurements. Since these operators do not commute with each other (except specific pairs of them), there is a non-trivial question if the code can or cannot operate with simultaneous continuous measurements. We have shown that such  operation is possible. An advantage of the continuous-measurement operation is that it requires only a passive steady-state monitoring of error syndromes, in contrast to repeated sequences of quantum gates between the code qubits and ancillary qubits  (followed by measurement of ancillas) to implement projective measurements.

Simultaneous measurement of non-commuting qubit operators \cite{Ruskov-2010} is a physically  interesting beyond-textbook process, which became an experimental reality only recently \cite{Hacohen-Courgy-2016}. Our work shows that it has relevance not only to foundations of quantum mechanics, but can also be useful for practical purposes, in this case for quantum error detection and correction.

The four-qubit Bacon-Shor code encodes one logical qubit, and the conventional operation involves random discrete evolution of an additional degree of freedom, the gauge qubit, due to sequential non-commuting projective measurements. In the continuous-measurement mode, the evolution of the gauge qubit becomes continuous, while transitions between the code space and error subspaces due to single-qubit errors remain similar to the projective-measurement case. As a result, the description of logical errors due to two close-in-time single-qubit errors remains somewhat similar in the continuous and projective measurement modes.

In the conventional Bacon-Shor code operation, the error syndrome is based on products (parity) of projective measurement results. In the continuous-measurement mode this is replaced by positive or negative signs of the cross-correlators between the noisy output signals; therefore the analysis relies on properties of correlators in continuous qubit measurements \cite{Korotkov-2001-spectrum, Korotkov-2011-correlation, Atalaya-2016-2}. Since the cross-correlators of noisy signals are very noisy, we need to construct time-averaged correlators; moreover, this averaging should involve at least two integrations over time. For the (inner) integration over the time difference in the two measurement channels, we used exponentially-decaying kernel and optimized over its time constant.
For the second (outer) integration over the running time, we considered two options: rectangular kernel and exponential kernel. Our results have shown that even though asymptotically the rectangular kernel is better, in the moderate range of parameters the exponential kernel is more natural.

The time constant $T_{\rm c}$ of the second integration is proportional to the response time $T_{\rm R}$: the delay between actual single-qubit error and obtaining an evidence that the error has occurred (crossing of a certain threshold by the time-averaged correlator). Since the logical error rate is proportional to $T_{\rm R}$, we would wish to decrease $T_{\rm c}$. However, this increases the rate of false alarms, when the error is mistakenly reported because of a large fluctuation of the time-averaged correlator. Therefore, there is a trade-off in the choice of $T_{\rm c}$ (Fig.\ \ref{fig:term-logical}).

A comparison between the code operations with projective and continuous measurements shows that they are comparable when the half-cycle duration $\Delta t$ of the projective-measurement mode is about $20\, \tau_{\rm m}$, where the strength of continuous measurement is characterized by the ``collapse'' (``measurement'') timescale $\tau_{\rm m}$. Even though this may seem to indicate that projective-measurement mode is easier to realize (allowing longer time scales), a comparable (if not larger) ratio $\Delta t/\tau_{\rm m}$ is implicitly assumed in the conventional operation with ``instantaneous'' projective measurements (when formally $\tau_{\rm m}=0$). As mentioned above, the advantage of the operation with continuous measurements is the absence of any time-dependent protocol (constantly repeated sequence of gates, ramping up and down measurement pulses, etc.).

Since the four-qubit Bacon-Shor code cannot perform quantum error correction and provides only quantum error detection, our results in this paper are formally applicable only to the quantum error detection with continuous measurement of non-commuting operators. We think that results for the nine-qubit Bacon-Shor code \cite{Poulin-2005,Bacon-2006,Terhal-2015} (which is an error correcting code) should in general be similar to the results in this paper; most importantly, we think that its operation with continuous measurement is really possible. The analysis can be based on evolution equation (\ref{evolution-rho}) with 12 measured gauge operators and four monitored time-averaged correlators constructed as three-signal products via Eq.\ (\ref{C-tilde-def-2}). However, we did not do any calculations for the nine-qubit code, and this analysis should be done in a separate paper.

While a simultaneous continuous measurement of non-commuting single-qubit operators has been already demonstrated \cite{Hacohen-Courgy-2016}, simultaneous measurement of non-commuting two-qubit operators has not been demonstrated, and so far there is no clear theoretical proposal for such a measurement. However, continuous quantum measurement of superconducting qubits is a rapidly developing field \cite{Vijay-2012, Hatridge-2013, Murch-2013, deLange-2014, Riste-2013, Roch-2014, Hacohen-Courgy-2016}, and we hope that the four-qubit Bacon-Shor code with continuous measurements analyzed in this paper can be realized experimentally reasonably soon.

\acknowledgments

The authors thank Andrew Jordan, Justin Dressel, Todd Brun, and Irfan Siddiqi for useful discussions.
J.A., M.B., and A.N.K. acknowledge support from ARO grants W911NF-15-1-0496 and W911NF-11-1-0268. L.P.P. acknowledges support from ARO grant W911NF-14-1-0272 and NSF grant PHY-1416578.


\end{document}